%% file: main.tex
\documentclass[aps,prc,floatfix,showpacs,twocolumn, preprintnumbers]{revtex4-1}
\usepackage{dcolumn}
\usepackage{bm}
\usepackage{color}
\usepackage{amssymb}
\usepackage{amsmath}
\usepackage{graphicx}
\usepackage{amsfonts}
\usepackage{slashed}
\usepackage{pstricks}
\usepackage{float}
\usepackage[colorlinks = true,
            linkcolor = blue,
            urlcolor  = blue,
            citecolor = blue,
            anchorcolor = blue]{hyperref}
\usepackage{array}
\allowdisplaybreaks
\usepackage{dcolumn}
\usepackage{epsf}
\usepackage{slashed}
\usepackage{tikz}
\usepackage{subfig}

\usetikzlibrary{shapes, arrows.meta, chains, scopes, positioning, matrix, calc, external}
\usepackage{color}
\definecolor{BrickRed}{rgb}{0.8, 0.25, 0.33}
\definecolor{gray}{rgb}{0.6,0.6,0.6}
\definecolor{darkgreen}{rgb}{0.0, 0.545098, 0.0}
\definecolor{mypink1}{rgb}{0.858, 0.188, 0.478}

\mathchardef\mhyphen="2D 

\begin{document}
\preprint{FERMILAB-PUB-20-368-T}
\title{A quantum Monte Carlo based approach to intranuclear cascades}
\author{
{Joshua} Isaacson $^{\, {\rm a} }$,
{William} I. Jay $^{\, {\rm a} }$,
{Alessandro} Lovato $^{\, {\rm b,c} }$,
{Pedro} A. N. Machado $^{\, {\rm a} }$,
{Noemi} Rocco $^{\, {\rm a,b} }$,
}
\affiliation{
$^{\,{\rm a}}$\mbox{Theoretical Physics Department, Fermi National Accelerator Laboratory, P.O. Box 500, Batavia, IL 60510, USA}\\
$^{\,{\rm b}}$\mbox{Physics Division, Argonne National Laboratory, Argonne, Illinois 60439, USA}\\
$^{\,{\rm c}}$\mbox{INFN-TIFPA Trento Institute of Fundamental Physics and Applications, Via Sommarive, 14, 38123 {Trento}, Italy}\\
}
\date{\today}

%
\date{\today}
\begin{abstract} 
We propose a novel approach to intranuclear cascades which takes as input quantum Monte Carlo nuclear configurations and uses a semi-classical, impact-parameter based algorithm to model the propagation of protons and neutrons in the nuclear medium.  
We successfully compare our simulations to available proton-carbon scattering data and nuclear-transparency measurements.
By analyzing the dependence of the simulated observables upon the ingredients entering our intranuclear cascade algorithm, we provide a quantitative understanding of their impact.
Particular emphasis is devoted to the role played by nuclear correlations, the Pauli exclusion principle, and interaction probability distributions.  
\end{abstract}
\maketitle
\section{Introduction}
The propagation of nucleons through the nuclear medium is an important aspect of nuclear reactions, heavy-ion collisions, and astrophysical environments. 
It is also crucial in the analysis of electron-nucleus scattering experiments 
(see e.g. Refs.~\cite{Barreau:1983ht, Makins:1994mm, Anghinolfi:1995bz, Abbott:1997bc, Dutta:2000sn, Dutta:2003yt, Rohe:2005vc,  Dai:2018xhi, Dai:2018gch, Murphy:2019wed}), neutrino oscillation experiments~\cite{Abe:2016tmq, Ayres:2007tu, Abe:2018uyc, Abi:2020wmh}, and dark matter searches~\cite{Akesson:2018vlm}.
For example, scattering of neutrinos on nuclei can produce a number of outgoing hadrons.
The multiplicity of the recoiling hadrons can indicate statistically if the incoming particle was more likely to be a neutrino or anti-neutrino~\cite{Palamara:2016uqu};
their momenta are correlated with the original neutrino energy and direction, which can be used to gather additional information, for instance, on the leptonic $CP$ phase 
from the sub-GeV atmospheric neutrino samples~\cite{Kelly:2019itm} or from searches for dark matter annihilation from the Sun~\cite{Rott:2016mzs, Rott:2019stu}.

The quantitative understanding of nucleons’ propagation in the nuclear medium would in principle require a fully quantum-mechanical description of the hadronic final state. 
Due to its tremendous difficulty, this problem has been tackled by introducing different approximations.  
The seminal papers by Serber and Metropolis et. al.~\cite{Serber:1947zza,Metropolis:1958sb,Bertini:1963zzc,Cugnon:1980zz} laid the foundations for the use of Monte Carlo techniques in semi-classical intranuclear cascades (INC) that assume classical propagation between consecutive scatterings. The latter are modeled using free-space elementary cross sections whose final state is modified to account for the Pauli principle. 
The results obtained by these first implementations of INC agree at least qualitatively with experimental data on nuclear transparencies and the frequency and angular distribution of emitted fast protons~\cite{Metropolis:1958sb}. 

Experiments in heavy-ion reactions have spurred theoretical efforts to describe the dynamical evolution of nucleus-nucleus collisions using transport methods~\cite{Bertsch:1984gb,Stoecker:1986ci, Bauer:1986zz, Bertsch:1988ik, Danielewicz:1991dh} based on the Kadanoff-Baym equations~\cite{KadanoffBaym:1962,Botermans:1990qi}.
The impossibility of solving these equations exactly on real-world computers  
requires the introduction of approximations, such as the gradient expansion leading to the Boltzmann-Uehling-Uhlenback equations (BUU). 
For instance, the Giessen Boltzmann-Uehling-Uhlenback (GiBUU) model is based on this truncated set of semi-classical kinetic equations, which describe the dynamics of the 
hadronic system explicitly in phase space and in time~\cite{Cassing:1990dr,Teis:1996kx}. 
It can be difficult to estimate systematic effects coming from the truncation of the transport equations.
Although initially developed to simulate heavy-ion collisions, GiBUU has been extended to the description of lepton and photon scattering on nuclei~\cite{Buss:2011mx,Mosel:2019vhx}. 

Over the years, several studies have been devoted to improving the accuracy and extending the predictive power of INC models, with a focus on the analysis of nuclear spallation processes, where a hadronic probe with energy from a few tens of MeV to a few hundred MeV strikes a nucleus~\cite{Cugnon:1996xf,Duarte:2007jd,Iwamoto:2010zzb,Sawada:2012hk}. 
In contrast to the standard Glauber approach~\cite{glauber1959lectures},  where the so-called frozen approximation is utilized, INC simulations explicitly account for the motion of the background particles or scattering centers~\cite{Golubeva:1997at}. 
The validity of the semi-classical propagation assumed in the INC has yet to be fully assessed.
However, genuine quantum-mechanical effects can either be safely neglected because they are expected to play a minor role---as for coherent scattering in nuclear transparency calculations---or they can be effectively parametrized by means of trajectory deflections and nuclear collective excitations, as in the analysis of nuclear spallation processes~\cite{Uozumi:2012fm}. 

Important examples of state-of-the-art INC codes used in hadron- and lepton-nucleus scattering analyses are the Li\`{e}ge INC~\cite{Boudard:2002yn},
NEUT~\cite{Hayato:2002sd}, nuance~\cite{Casper:2002sd}, PEANUT (used within FLUKA)~\cite{Battistoni:2013tra,Battistoni:2015epi}, NuWro~\cite{Golan:2012wx}, and GENIE~\cite{Andreopoulos:2009rq} programs.
Despite some differences in technical aspects and degree of sophistication, all the above INC models use as input elementary cross sections and mean-field properties of nuclei, such as single-nucleon densities. 
A notable exception is NuWro, which has recently been extended to incorporate effective nucleon-nucleon correlations~\cite{Niewczas:2019fro}. 

In this work, we propose a novel cascade model that employs
nuclear configurations obtained from quantum Monte Carlo (QMC) calculations, which retains all correlation effects. 
The probability that the propagating nucleon scatters with a background particle is
modeled using two different weight functions that depend on the impact parameter of the two nucleons and their cross section. 
The nucleon-nucleon cross sections used in the simulation are taken to be the same as in  vacuum; in-medium effects are partially accounted for by imposing the Pauli exclusion principle and an effective nuclear binding.

Our semi-classical approach allows for the calculation of exclusive quantities in nuclear scattering, such as the fully differential phase space and the number of outgoing nucleons.
To test the validity of our model, we compare our simulations with available data on proton-nucleus scattering cross sections and nuclear transparencies.

On the technical side, the INC code developed here is publicly available at \url{https://github.com/jxi24/IntranuclearCascade}.
The front-end is written in \texttt{Python 3}, while computationally intensive code is written in \texttt{C++}, using \texttt{pybind11} to generate the interface code.
The source code is organized to be portable and modular, making it easy to use, extend, and improve.
All validations are available in the source code and can be easily performed by the front-end user.

\section{The physics of the impact parameter based intranuclear cascade model}

We now discuss in detail the physical content of our INC and how several ingredients were incorporated in the model.
The INC starts by generating a nuclear configuration, which describes the spatial distribution of protons and neutrons inside the nucleus. A large number of these configurations are computed from quantum Monte Carlo (QMC) methods~\cite{Carlson:2014vla} that use as input the highly-realistic Argonne $v_{18}$ (AV18)~\cite{Wiringa:1994wb} plus Illinois 7 (IL7)~\cite{Pieper:2008} Hamiltonian -- more details can be found in Sec.~\ref{sec:nuclear-configuration}. All particles in a nuclear configuration are initially labeled as background particles. After randomly picking a nuclear configuration, a random nucleon is selected, struck, and labeled as a propagating nucleon. 
Propagating nucleons are assumed to be  point-like, on-shell particles moving at some velocity given by their four-momentum. 
The spatial coordinates of background nucleons are kept fixed until they interact with a propagating particle. In that case, their momentum is sampled from either a local or global Fermi gas distribution.

\begin{figure}[t]
    \centering
    \includegraphics[width=.45\textwidth]{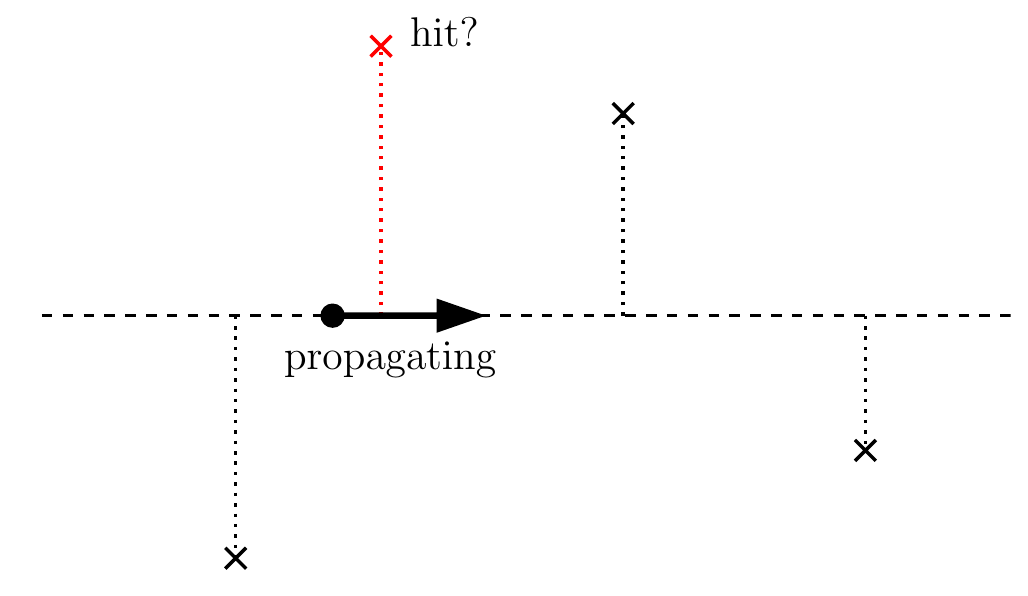}
    \caption{Schematics of a propagating nucleon (black circle). The distance travelled by the propagating nucleon is depicted by a black arrow. The crosses represent background nucleons. An interaction test would be performed for the nucleon in red. 
    \label{fig:cylinder}}
\end{figure}

The system is evolved in time steps, which is a parameter of the cascade model, and should be chosen small enough to simulate the cascade accurately but large enough to run in a reasonable amount of time. At each time step, the following procedure is performed for each propagating nucleon.
First, it is checked if there are any background particles within the volume perpendicular to the line segment defined by the initial to the final position of a propagating particle; this is schematically shown in Fig.~\ref{fig:cylinder}. 
The impact parameter is calculated for all such nucleons, and this list is sorted from closest to furthest.
Then an accept-reject step is performed on this list, until either an interaction takes place or the end of the list is reached.
This test determines if an interaction occurred according to a probability distribution.
If an interaction occurred, the phase space is generated using a fully differential nucleon-nucleon cross section.
The Pauli exclusion principle is approximately taken into account by comparing the magnitude of the momentum of the final state particles with the Fermi momentum $k_F$ (the user can choose between either the global or the local values). The interaction takes place only if both momenta are above $k_F$.
Otherwise, the interaction does not happen, and the propagating particle keeps its original four momentum.

If the interaction took place, the outgoing particles are both treated as propagating particles, and a formation zone is set for each of them~\cite{Landau:1953gr, Stodolsky:1975qq} (see also Ref.~\cite{Golan:2012wx}).
Note that, as a distinctive feature of our INC model, the interaction is finite ranged and the outgoing nucleons do not have to be at the same position in space.
Finally, a propagating particle that reaches a sufficiently large radius may exit the nucleus if its kinetic energy overcomes the effective nuclear binding.
If there is insufficient energy to escape, the particle is relabeled as a background particle.
Otherwise, this effective binding energy is  subtracted from the particle's energy, and then the particle is labeled as a final particle and stops propagating.
The INC stops when there are no more propagating particles in the nucleus.

The structure of the algorithm is summarized in Fig.~\ref{fig:algorithm}.
In what follows, we provide the details and expressions that are used in our impact parameter based INC. 

\input{tikz/Algorithm.tex}

\subsection{Nuclear configuration}\label{sec:nuclear-configuration}
Nuclei are complicated many-body systems of fermions, whose structure and dynamics emerge from individual interactions among the constituent protons and neutrons. To a remarkably large extent, the latter can be modeled by the non-relativistic Hamiltonian 
\begin{equation}
H=\sum_{i}\frac{{\bf p}_i^2}{2m_N}+\sum_{i<j} v_{ij}+ \sum_{i<j<k}V_{ijk}\,,
\label{NMBT:ham}
\end{equation}
where ${\bf p}_i$ denotes the momentum of the $i$-th nucleon with mass $m_N$, while $v_{ij}$ and $V_{ijk}$ are the nucleon-nucleon ($N\!N$) and three-nucleon ($3N$) potentials, respectively.
We employ the highly-realistic Argonne $v_{18}$ (AV18) $N\!N$ interaction~\cite{Wiringa:1994wb} that includes spin, isospin, tensor, spin-orbit and quadratic momentum-dependent terms as well as isospin-symmetry-breaking corrections and reproduces the Nijmegen nucleon-nucleon database with a $\chi^2$/datum $\simeq 1$. 
For the $3N$ interaction, we use the Illinois-7 (IL7) potential~\cite{Pieper:2008}, which consists of the dominant standard Fujita-Miyazawa two-pion exchange and smaller multi-pion-exchange components resulting from the excitation of intermediate $\Delta$ resonances.
The IL7 potential also contains phenomenological isospin-dependent central terms.
The parameters characterizing this three-body potential have been determined by fitting the low-lying spectra of nuclei in the mass range $A$=3--10.
The overall AV18+IL7 Hamiltonian then leads to predictions of $\approx 100$ ground- and excited-state energies up to $A$=12, including the $^{12}$C ground- and Hoyle-state energies, in good agreement with the corresponding empirical values~\cite{Carlson:2014vla}.

Over the last decades significant progress has been made in the development of ab initio nuclear methods, which solve the many-body Schr\"odinger equation associated with the Hamiltonian of Eq.~\eqref{NMBT:ham} with controlled approximations~\cite{Bogner:2009bt,Barrett:2013nh,Hagen:2013nca,Barbieri:2016uib}.
For light nuclei, quantum Monte Carlo (QMC) and, in particular, Green's function Monte Carlo (GFMC) methods have been exploited to carry out calculations of nuclear properties, based on realistic $N\!N$ and $3N$ potentials, and consistent one- and two-body meson-exchange currents~\cite{Carlson:2014vla}.  
GFMC begins with the construction of a trial wave function $\Psi_T$ that is a symmetrized product of two- and three-body correlation operators acting on an antisymmetric $A$-body single-particle wave function that has the proper quantum numbers for the state of interest.
The variational parameters in $\Psi_T$ are found by minimizing the energy expectation value 
\begin{equation}
E_0 \leq E_T = \frac{\langle \Psi_T | H | \Psi_T \rangle} {\langle \Psi_T | \Psi_T \rangle}\, ,
\end{equation}
where $E_0$ is the true ground-state energy of the system. The calculation of $E_T$ requires the numerical solution of a multidimensional integral that is carried out employing standard Metropolis Monte Carlo sampling in configuration space.

GFMC then projects out the lowest eigenstate $\Psi_0$ of the given quantum numbers starting from $\Psi_T$ by performing a propagation in imaginary time $\tau$
\begin{equation}
|\Psi_0\rangle = \lim_{\tau \rightarrow \infty} {\rm exp}[-(H-E_0)\tau] |\Psi_T\rangle.
\end{equation}
The propagation $|\Psi(\tau)\rangle = {\rm exp}[-(H-E_0)\tau] |\Psi_T\rangle$ is carried out as a series of many small imaginary-time steps $\Delta\tau$. Expectation values of operators are evaluated as mixed matrix elements $O(\tau) = \langle \Psi_T | O | \Psi(\tau)\rangle$, and the behavior as a function of $\tau$ analyzed to obtain converged results.
Because $H$ and ${\rm exp}[-(H-E_0)\tau]$ commute, the mixed estimate is the exact expectation of $\langle \Psi(\tau/2) | O | \Psi(\tau/2)\rangle$ but linear extrapolations are used to evaluate other quantities. 

In addition to binding energies the GFMC provides detailed information on the distribution of nucleons in a nucleus in both coordinate and momentum space, which are interesting in multiple experimental settings. For example, the mixed-estimate of the single-nucleon density is calculated as
\begin{align}
	\rho_{N}(r) &=\frac{1}{4\pi r^2}\big\langle\Psi_T \big|\sum_i \mathcal \delta(r-|\mathbf{r}_i|)P_{N}\big|\Psi(\tau)\big\rangle\,, 
	\label{eq:rho_N}
\end{align}
where $N=p,n$; $P_{N_i}=\frac{1\pm\tau_{z_i}}{2}$ is the neutron or proton projector operator; and, $\rho_N$ integrates to the number of protons or neutrons. The two-body density distribution, yielding the probability of finding
two nucleons with separation $r$, is defined as
\begin{align}
	\rho_{NN}(r) &=\frac{1}{4\pi r^2}\big\langle\Psi_T \big|\sum_{i<j} \mathcal \delta(r-|\mathbf{r}_{ij}|)P_{N_i} P_{N_j} \big|\Psi(\tau)\big\rangle\,.
	\label{eq:rho_NN}
\end{align}

\begin{figure}[t]
    \centering
\includegraphics[width=.53\textwidth]{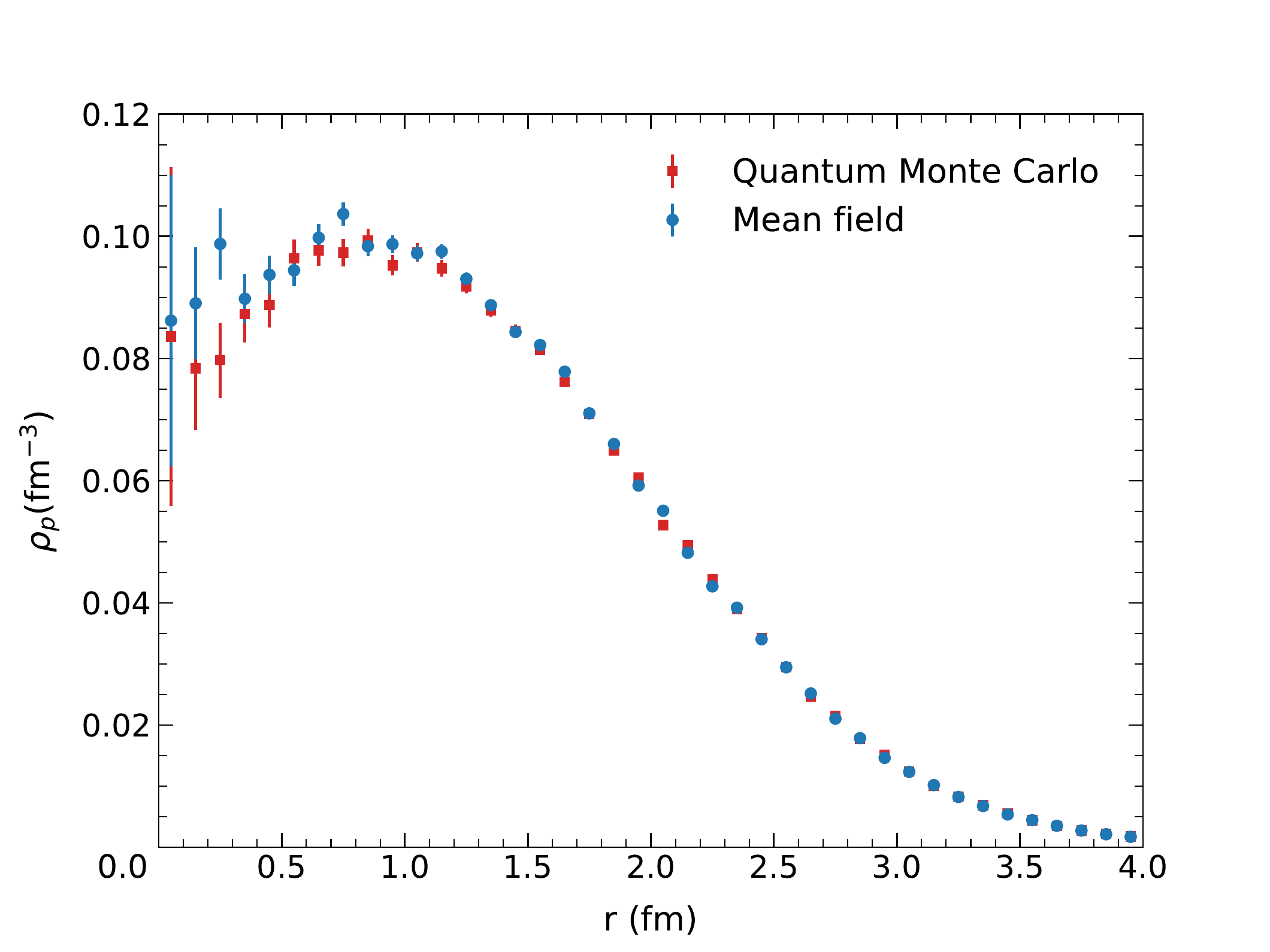}
    \caption{Nucleon density in carbon from Green's function Monte Carlo (red) and mean field (blue) configurations.
    \label{fig:density_1b}}
\end{figure}

The positions of the constituents protons and neutrons utilized in the nuclear cascade algorithm are sampled from $36000$ GFMC configurations. We employ the so-called constrained-path approximation~\cite{Wiringa:2000gb} to make sure that their Monte Carlo weights remain positive, thereby facilitating their usage in the cascade algorithm. As a consequence, the single-proton distribution displayed by the blue solid circles of Fig.~\ref{fig:density_1b} is slightly different from the results reported in Ref.~\cite{Lovato:2013cua}, which have been obtained performing fully unconstrained imaginary-time propagations. Since we neglect the charge-symmetry breaking terms in the Hamiltonian, and since $^{12}$C is isospin symmetric, the single-neutron distribution is identical to that of the proton. 

\begin{figure}[h]
    \centering
\includegraphics[width=.53\textwidth]{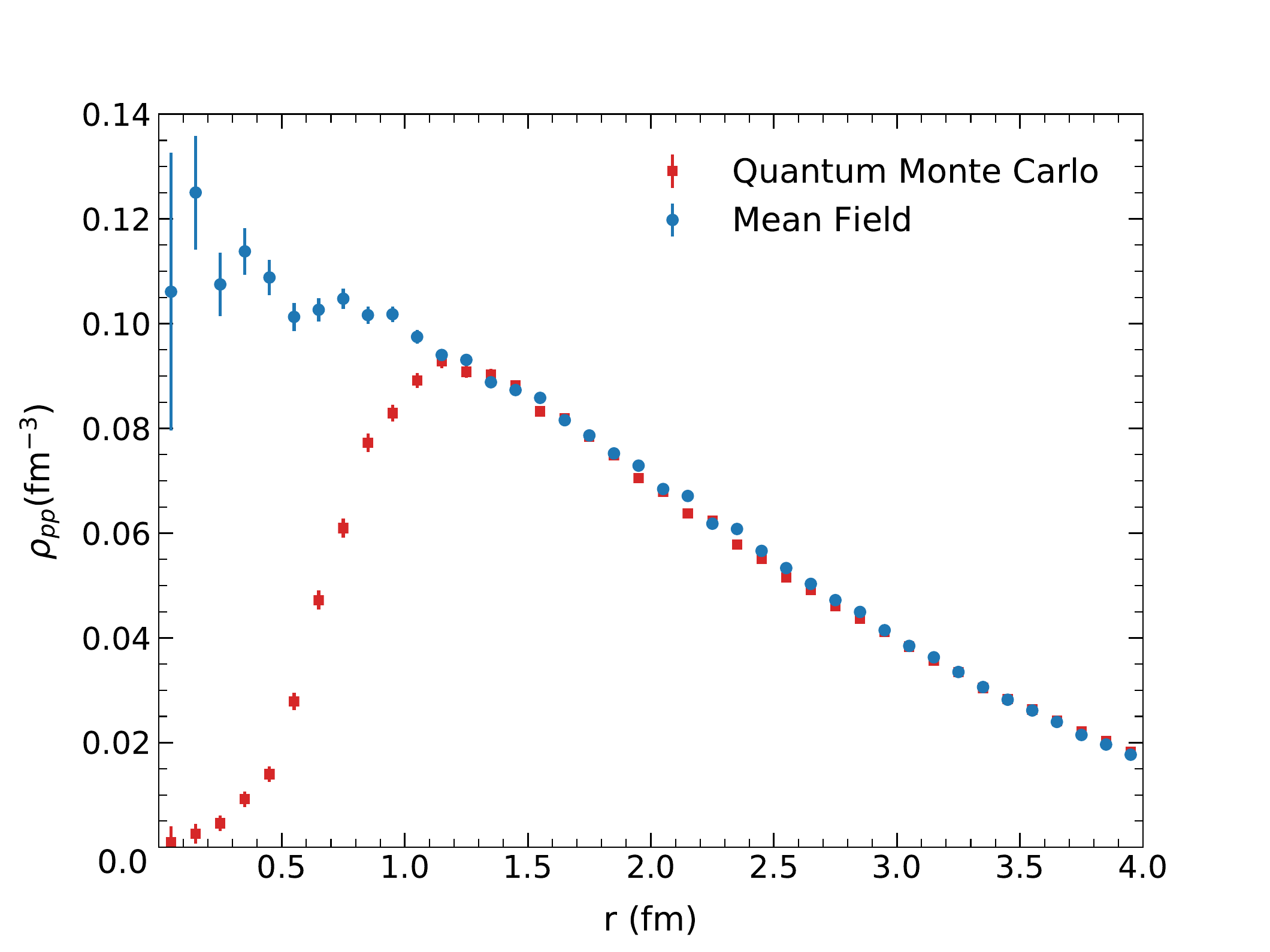}
\includegraphics[width=.53\textwidth]{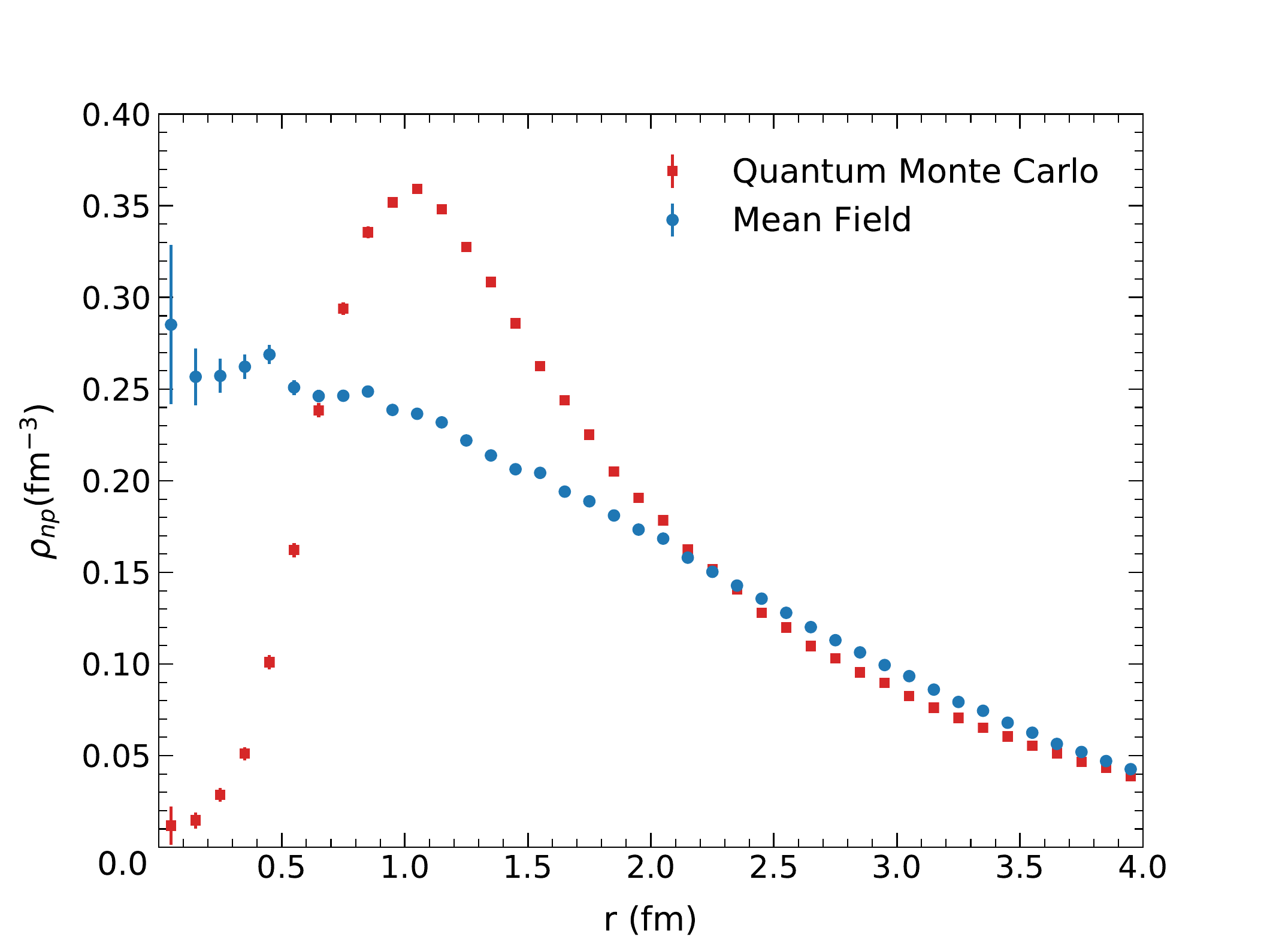}
    \caption{Proton-proton (top panel) and proton-neutron (bottom panel) correlation functions in carbon from Green's function Monte Carlo (red) and mean field (blue) configurations.
    \label{fig:density_12b}}
\end{figure}

For benchmark purposes, we also sample $36000$ mean-field (MF) configurations from the single-proton distribution.
The corresponding single-proton densities coincide by construction with the GFMC one, as shown in Fig.~\ref{fig:density_1b}.
However, the differences between GFMC and MF configurations become apparent when comparing the corresponding two-body density distributions represented in Fig.~\ref{fig:density_12b}.
The short-range repulsive core of the $NN$ interaction prevents two nucleons from being close to each other.
As a consequence, the $pp$ and $np$ GFMC density distributions are small at short separation distances.
Furthermore, the difference between the GFMC $pp$ and $np$ density distributions around $r=1$ fm can be attributed to the strong tensor correlations induced by the one-pion-exchange part of the $NN$ interaction, which is further enhanced by the two-pion-exchange part of the $3N$ potential.
Note that the short-range behavior of $\rho_{NN}$, which is largely nucleus independent, does depend strongly on the $NN$ interaction model~\cite{Cruz-Torres:2019fum}.
On the other hand, the MF ones do not exhibit this rich behavior as the correlations among nucleons are entirely disregarded. 

\subsection{Nucleon momentum distribution} 
As mentioned above, when a nucleon is struck, its momentum is obtained assuming either a local or global Fermi gas distribution. 
In the case of the local Fermi gas,
the magnitude of the three-momentum is randomly sampled in the interval $[0,k^N_F(r)]$ where $k^N_F(r)$ is the Fermi Momentum defined in terms of the single nucleon density $k^N_F(r)=(\rho_N(r)3\pi^3)^{1/3}$ and $N=p,n$. 
In the case of the global Fermi gas, the momentum is determined in the same way, but $k^N_F$ is position independent. 
The local Fermi gas model is known to provide a more realistic nucleon momentum distribution for finite nuclei than the global Fermi gas. 
For this reason, although both models are implemented in our code, we only present results based on the local Fermi gas predictions.
In the future, we plan to include more accurate nucleon momentum distribution, based on state-of-the-art many-body calculations that properly account for nuclear correlations. 

\subsection{Nucleon-nucleon interaction algorithm}
To check if an interaction between nucleons occurs, an accept-reject test is performed on the closest nucleon according to a probability distribution $P(b)$ (see e.g. Ref.~\cite{Sjostrand:1987su} for similar considerations)
where 
$b$ is the impact parameter. 
We impose two conditions on this probability,  
\begin{equation}\label{eq:int_prob_conds}
    P(0)=1\quad\textrm{and}\quad
    \int_0^{2\pi}\int_0^\infty d\varphi \,b db P(b) = \sigma,
\end{equation}
where the cross section $\sigma$ depends on the incoming particle content and the center-of-mass energy, which is sampled from the nuclear configuration.
The second condition ensures that the mean free path of a nucleon traveling in a medium of uniform density is $\lambda_{\rm mfp} = 1/\sigma \bar{\rho}$, where $\bar{\rho}$ is the number density.

Two implementations of $P(b)$ have been studied here. The first we dub the \emph{cylinder} interaction probability, 
\begin{equation}
   P_{\rm cyl}(b) = \Theta(\sigma/\pi-b^2),
\label{prob:cyl}   
\end{equation}
where $\Theta(x)=1$ if $x\ge 0$, else $\Theta(x)=0$. 
This probability mimics a more classical, billiard ball like system, where each billiard ball has a radius $\approx \sqrt{\sigma/\pi}$.
The second implementation is the \emph{Gaussian} interaction probability
\begin{equation}
   P_{\rm Gau}(b) \equiv \exp\left(- \frac{\pi b^2}{\sigma}\right),
\label{prob:gauss} 
\end{equation}
which is inspired by the work of Ref.~\cite{Sjostrand:1987su}.
Both $P_{\rm cyl}$ and $P_{\rm Gau}$ satisfy the conditions in Eq.~\eqref{eq:int_prob_conds}.
We use the nucleon-nucleon cross sections from the SAID database~\cite{SAID} obtained using \texttt{GEANT4}~\cite{Agostinelli:2002hh}, or from the NASA parametrization~\cite{Norbury:2008}.

\subsection{Phase space, Pauli blocking and after-interaction}
If an interaction occurred, the phase space of the outgoing particles is generated using  fully differential nucleon-nucleon cross sections.
Note that, at the moment, we only include protons and neutrons in our INC model.
Pauli blocking enforces Fermi-Dirac statistics for the nucleons and amounts to testing whether their final-state momenta are above the Fermi momentum. 
Two different models of the Pauli exclusion principle have been approximately implemented. 
The \emph{global} and \emph{local} Pauli blocking routines essentially forbid a scattering if the momentum of any of the final state particles is below the average Fermi momentum (for the global Fermi gas model) or the local Fermi momentum (for the local Fermi gas model), respectively.
We emphasize again that, although we have implemented the global Fermi gas model, we do not report any results using it.

If the interaction took place, the outgoing particles are both treated as propagating particles, and a formation zone is set for each of them~\cite{Landau:1953gr, Stodolsky:1975qq} (see also Ref.~\cite{Golan:2012wx}).
The formation zone is a length or period of time in which a particle does not interact with any nucleons.
This models the coherence and interference of interactions in quantum mechanics.
The formation zone is given by
\begin{equation}
   \delta t = \frac{E'}{m_N^2-p\cdot p'},
\end{equation}
where $m_N=938$~MeV is the nucleon mass, $E'$ is the energy of the outgoing nucleon, and $p$ and $p'$ are the four momenta of the  incoming and outgoing nucleon respectively. 

\subsection{Exiting the nucleus}
When the radial position of a propagating particle is larger than a certain distance, much larger than the nuclear radius, a test is performed to check if the particle has enough energy to escape the nuclear potential.
If its kinetic energy is larger than the the nuclear potential barrier, then the particle escapes the nucleus and is labeled as a final state particle. The momentum of the particle is modified to include the effective nuclear binding. Final state particles do not propagate and are essentially the input that should be given to a detector simulation.

\section{Results and Model Validation}
In this Section, we present the validation tests we performed and the results we obtained within our INC model.
For comparison purposes, we also implement a version of the nucleon-nucleon interaction algorithm that we dub \emph{mean free path} (MFP). 
This approach is routinely used in event generators. 
Within this algorithm, the system is not evolved in time steps but rather in constant position steps that we indicate as $\lambda_{\rm max}$. 
At each step of the loop the mean free path of the propagating particle reads
\begin{equation}
    \tilde{\lambda}=1/\Big(\rho_{p}(r)\sigma_{Np}+\rho_{n}(r)\sigma_{Nn}\Big)
\end{equation}
where $N=n,p$ refers to the isospin of the propagating particle and $r$ its distance from the centre of the nucleus. 
Note that in the limit of constant density, $\rho_{p(n)}(r)=\bar{\rho}$, we would recover $\tilde{\lambda}=\lambda_{\rm mfp}$.
(see also Sec.~\ref{ssec:mean_free_path} below).
The probability that the struck particle traveled a distance $\lambda$ without interacting can be written as
$P(\lambda)=e^{-\lambda/\tilde{\lambda}}$.
We can then sample $\lambda =-\tilde{\lambda}\cdot \ln({\rm rand}[0,1])$ and say that an interaction took place if $\lambda < \lambda_{\rm max}$. Note that $\lambda_{\rm max}$ has to be chosen small enough in order to  
satisfy the assumption of a constant density.

\begin{figure}[t]
    \centering
    \includegraphics[width=.53\textwidth]{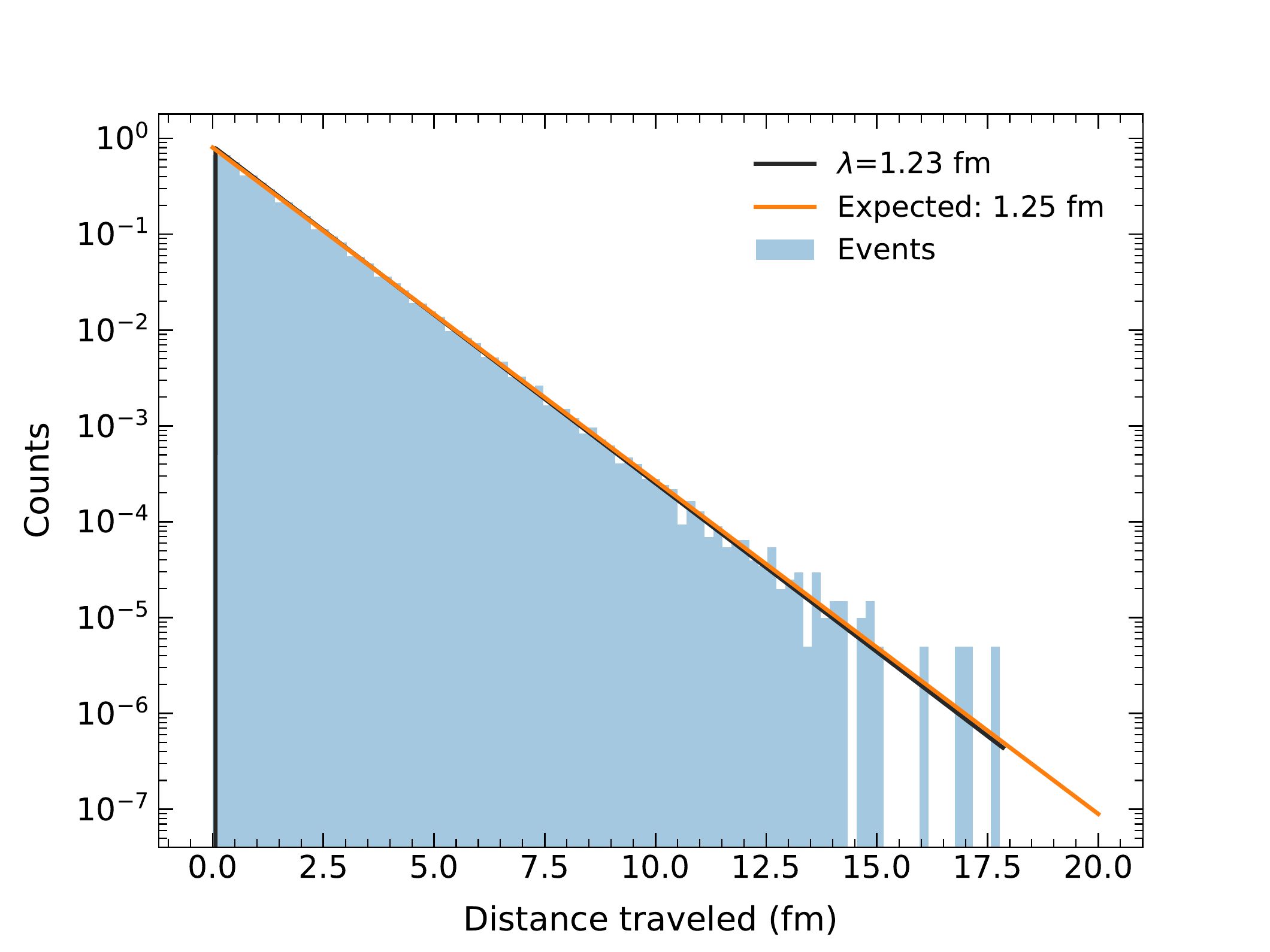} \\
    \caption{Distribution of distance traveled (blue histogram) by 500 MeV test nucleons traveling through a $0.16$ nucleons/fm$^3$ background density of nucleons at rest, fixed interaction cross-section of 50~mb, and using the Gaussian interaction probability.
    The black and orange lines are the expected distributions for the fitted (1.23~fm) and expected (1.25~fm) values of the mean free path, respectively.
    \label{fig:mfp}}
\end{figure}

\subsection{Mean free path \label{ssec:mean_free_path}}
The simplest validation consists of calculating the mean free path of a nucleon travelling through a medium of uniformly distributed nucleons, with fixed interaction cross sections.
Our goal is to obtain 
\begin{equation}\label{eq:mfp}
    \lambda_{\rm mfp} = \frac{1}{\sigma \rho},
\end{equation}
where $\sigma$ is the nucleon-nucleon cross section, and $\rho$ is the number density of nucleons.
We simulate a background of static, randomly and uniformly distributed protons and neutrons at rest.
For an event, we insert into the medium a test particle with a fixed energy and propagate it through the medium.
When the first interaction happens, we record the length the test particle has traveled.
Figure~\ref{fig:mfp} shows the distribution of lengths traveled for a 500 MeV test proton in a $0.16$ nucleons/fm$^3$ background density of nucleons, assuming a fixed interaction cross section of 50~mb, and using the Gaussian interaction probability. 
The expected distributions, according to Eq.~\eqref{eq:mfp}, are also shown for the fitted (black line) and the expected (orange line) values of the mean free path. 
The nuclear density and interaction cross section were chosen arbitrarily. Adjusting these values does not change the agreement between the expected mean free path and the calculated mean free path. 
While not shown here, using the cylinder interaction probability does not change the results either.
As we can see, our code correctly reproduces the expected behavior for the mean free path, allowing us to proceed to more complex tests of our INC.

\subsection{Proton-carbon Scattering Data}
Reproducing the proton-nucleus   
cross section measurements is an important 
test of the accuracy of the INC model. 
Proton-nucleus scattering probes the nucleon-nucleon cross section which is
 typically divided into two pieces, the reaction and the elastic cross sections,
\begin{equation}
    \sigma_{\rm tot} = \sigma_R + \sigma_{\rm el}.
\end{equation}
In the elastic part, no energy is transferred into nuclear excitation and the nucleus remains unbroken, that is $n+A \rightarrow n+A$.
The reaction cross section includes transition to nuclear excited states, $n+A \rightarrow n+A^*$, as well as inelastic reactions $n+A \rightarrow X$.

Several experiments have been carried out to determine the total reaction cross section, see for example Refs.~\cite{Dicello:1970mx, Bobchenko:1979hp, Bauhoff:1986gcb, Kox:1987qw, Sihver:1993pc, Carlson:1996ofz}.
The latter is typically obtained by measuring the total cross section from the change in intensity of a calibrated proton beam traversing a carbon target and then subtracting the calculated elastic cross section.

\begin{figure}[ht]
    \centering
    \includegraphics[width=.5\textwidth]{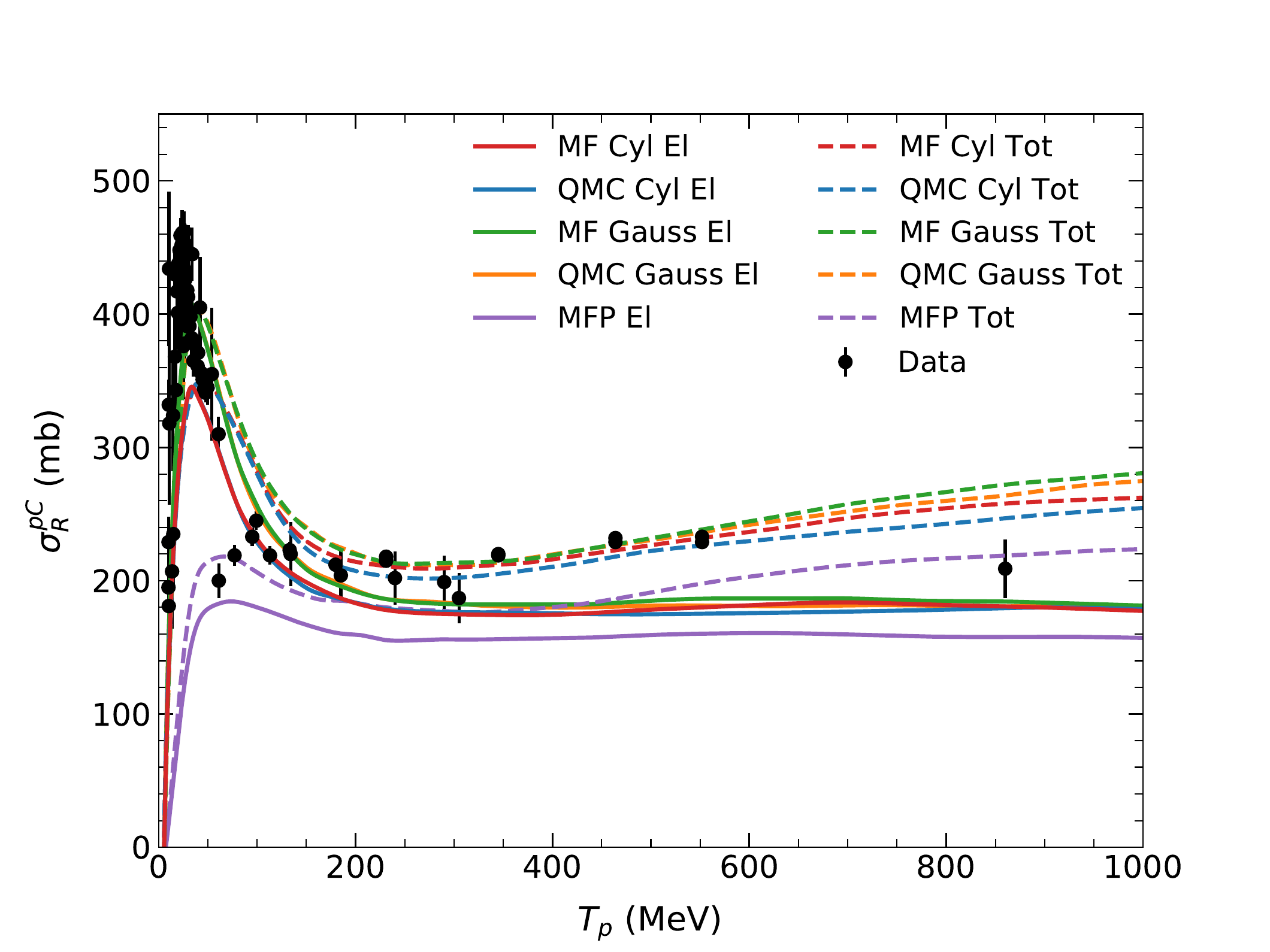}\\
    \includegraphics[width=.5\textwidth]{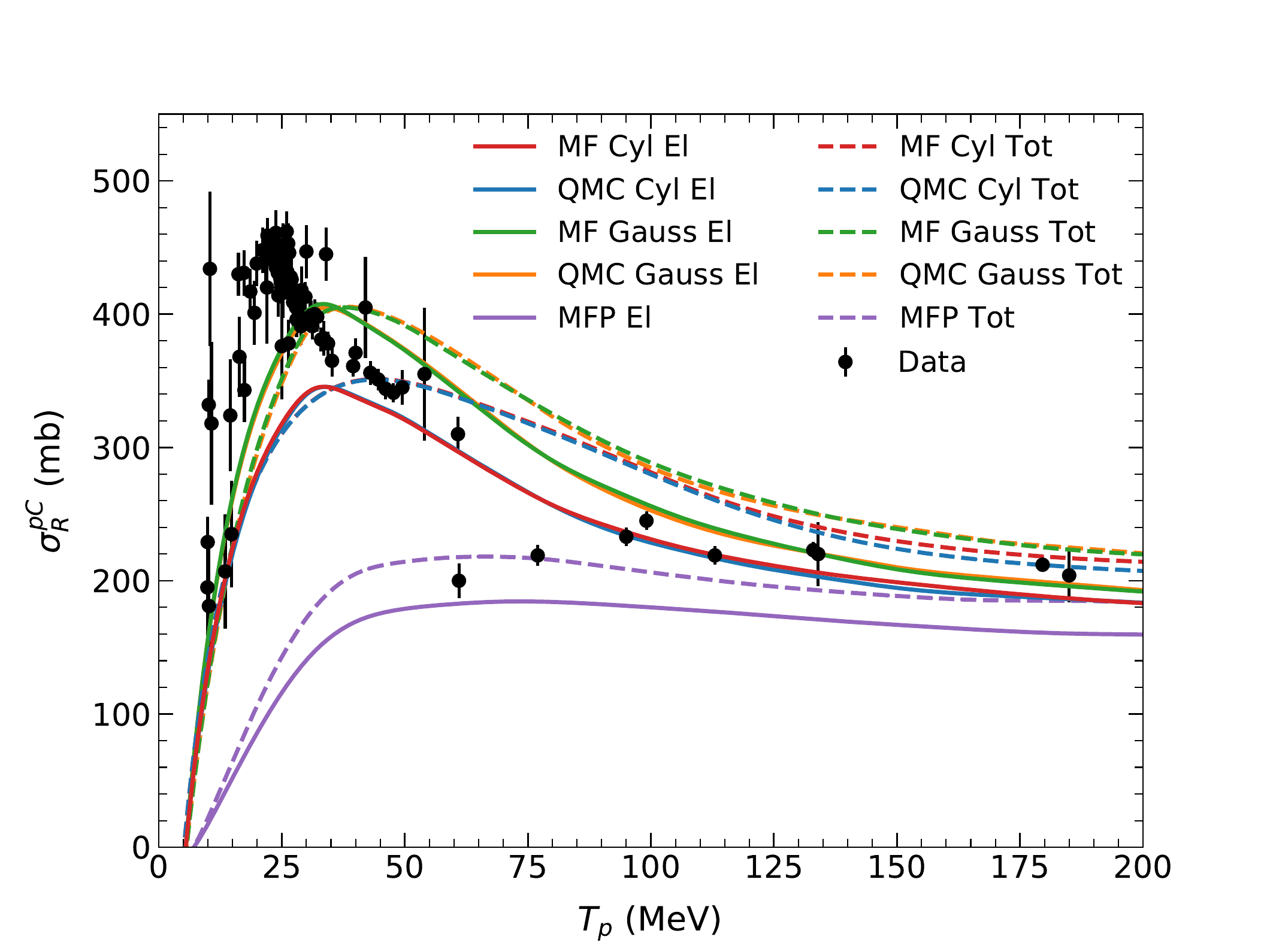}\\
    \caption{Proton-carbon scattering total cross section as a function of the incoming proton kinetic energy. 
    In the upper panel the entire energy range for which experimental data are available is shown. In the lower panel the low energy region is magnified. 
    The red and blue curves correspond to the cylinder algorithm where the mean field (MF) and quantum Monte Carlo (QMC) configurations have been used, respectively. The green and orange curves are the same but for the Gaussian interaction probability. The results displayed in purple refers to the mean free path (MFP) calculations.
    The solid and dashed curves corresponds to the use of the \texttt{GEANT4}~\cite{Agostinelli:2002hh} and NASA~\cite{Norbury:2008} parametrization of the cross section in the interaction probability, respectively. 
    The data points are from Ref.~\cite{Carlson:1996ofz}
    \label{fig:pC_xsec}}
\end{figure}

We compute $\sigma_R$ neglecting Coulomb interactions, as they are expected to  contribute mostly to $\sigma_{\rm el}$. 
We obtain the proton-carbon scattering cross section by the following simulation (with a different setup from the proposed algorithm of Fig.~\ref{fig:algorithm}).
We define a beam of protons with energy $E$, uniformly distributed over an area $A$ (orthogonal to the proton momenta).
Note that $A\gg \pi R^2$, where $R$ is the radius of the carbon nucleus.
The carbon nucleus is situated in the center of the beam.
We propagate each proton in time and check for scattering at each step. 
The Monte Carlo reaction cross section is then defined as the area of the beam times the fraction of scattered events, namely,
\begin{equation}
    \sigma_{\rm MC}=A \frac{N_{\rm scat}}{N_{\rm tot}}.
\end{equation}
This is not exactly the experimentally measured reaction cross section.
Angular and/or momentum acceptances for the attenuated beam are finite, and we do not include these effects in our calculation.
Nevertheless, we do not expect such effects to  change our results significantly, and thus $\sigma_{\rm MC}$ should be a good approximation of the reaction cross section. 
Moreover, imposing Pauli blocking on both outgoing nucleons will effectively suppress the contribution of elastic transitions.

The two panels of Fig.~\ref{fig:pC_xsec} display the proton-carbon scattering cross sections as a function of the proton kinetic energy. In the upper panel our Monte Carlo simulations are compared with experimental data in the entire energy region in which data are available~\cite{Carlson:1996ofz}, while the lower panel focuses on proton kinetic energies below 200~MeV. 
The curves correspond to different implementations of the INC.
These implementations are composed of three ingredients, namely,
\begin{enumerate}
   \item Nuclear configuration: quantum Monte Carlo (QMC) or mean field (MF);
   \item Interaction model: cylinder (cyl), Gaussian (Gauss), or mean free path (MFP);
   \item Nucleon-nucleon cross section: elastic (El) or total (Tot).
\end{enumerate}
Note that the interaction model MFP does not use any configuration, but rather the local density.

The solid lines have been obtained using the nucleon-nucleon cross sections from the SAID database~\cite{SAID}, obtained using \texttt{GEANT4}~\cite{Agostinelli:2002hh}, in which only the elastic contribution is retained. 
The dashed lines used the NASA parameterization~\cite{Norbury:2008}, which includes inelasticities.
The inelastic contribution leads to an enhancement of the cross section which is necessary to reproduce the data correctly for intermediate to large proton kinetic energies.
The NASA parameterization allows us to approximate the effects of including pions in the cascade.
Results obtained with these two nucleon-nucleon cross sections are indicated in Fig.~\ref{fig:pC_xsec} by ``El'' for elastic cross section and ``Tot'' for the total cross section.
As expected, the $p$-carbon cross section obtained using only the elastic nucleon-nucleon cross section is consistently lower than the one obtained using the total nucleon-nucleon cross section.
The effect is large for $T_p>50$~MeV, where pion production becomes relevant.
In a future work, we will explicitly include pion degrees of freedom in our INC as these are known to play a crucial role in this region.
Additionally, the present results neglect modifications to the nucleon-nucleon cross section arising from the nuclear medium itself.
Such modifications have been considered in the literature ~\cite{Pandharipande:1992zz}.
We also leave these considerations for a future work.

The dependence of the results on the functional form of the interaction probability and on the nuclear model adopted to generate nuclear configurations has also been investigated.  
The blue and red lines are obtained from the cylinder probability of Eq.~\eqref{prob:cyl} and sampling the initial nucleon configurations from the quantum Monte Carlo (QMC) and mean field (MF) distributions, respectively. To obtain the orange and green curves, QMC and MF configurations were used together with the Gaussian probability of Eq.~\eqref{prob:gauss}. We observe that in both cases, the QMC and MF curves are almost superimposed, indicating that this observable does not depend strongly on correlation effects among the nucleons. A more detailed discussion of their impact on particle propagation distances will be discussed in Sec.~\ref{sec:corr:eff}. 

The purple lines refer to mean free path (MFP) calculations which present conceptual differences with respect to the cylinder and Gaussian cases and yield large discrepancies in the results. The predictions for the $p$-carbon cross section obtained from the MFP approach underestimate the experimental data and are significantly lower with respect to the other curves obtained with more refined techniques.

In the lower panel of Fig.~\ref{fig:pC_xsec}
we focus on the comparison between the different MC simulations and the data in the low energy region, i.e., proton kinetic energies below $200$ MeV.
The results obtained using the Gaussian and cylinder algorithm are in fair agreement with the experimental data: the curves display the correct behaviour although the position of the peak is not exactly reproduced.
We note that Pauli blocking plays a fundamental role for these kinematics, significantly quenching the results.
More sophisticated techniques aimed at consistently orthogonalizing the nuclear wave function in the final state are discussed in Ref.~\cite{Gonzalez-Jimenez:2019qhq}.
Their implementation in our INC will be the subject of a future work. 

The functional forms adopted to determine the probability distribution (Gaussian vs. cylinder) 
yield predictions which differ by $\sim15\%$ in the low energy region (compare, for instance, the green and red lines in Fig.~\ref{fig:pC_xsec}).
For intermediate proton kinetic energies $T_p$ the curves converge to the same result and reproduce the data.
At low $T_p$, the predicted cross sections for the MFP method are $\sim50\%$ smaller than those of the Gaussian and cylinder methods.
As these differences are also present in nuclear transparency, we will discuss their origins in the next section.

\subsection{Nuclear transparency}

The nuclear transparency yields the average probability that a struck nucleon (either a proton or a neutron) leaves the nucleus without interacting with the spectator particles. 
Measurements of the nuclear transparency to high energy protons in quasielastic $e+A\to e+p+A'$  scattering have been carried out by a number of experiments ~\cite{ONeill:1994znv, Abbott:1997bc, Rohe:2005vc, Garrow:2001di, Dutta:2003yt, Garino:1992ca}.
These measurements are performed with fixed kinematics (i.e., for a fixed incoming beam energy and scattering angle) and measure the outgoing electron and proton.
The proton produced in the primary vertex can be absorbed or deflected while exiting the nuclear medium because of final state interactions with the remnant system, leading to a reduction of the measured $e+A\to e+p+A'$ cross section.
This reduction defines the nuclear transparency, which is given by the ratio of the observed events to events predicted in the Plane Wave Impulse Approximation (PWIA) 
\begin{equation}
    T = \frac{N_{\rm exp}}{N_{\rm PWIA}}.
\end{equation}
To obtain $N_{\rm PWIA}$ the initial proton is treated as a bound state with energy and momentum distribution described by a spectral function; while the final proton is a free particle state propagating as a plane wave.

\begin{figure}[t]
    \centering
    \includegraphics[width=.5\textwidth]{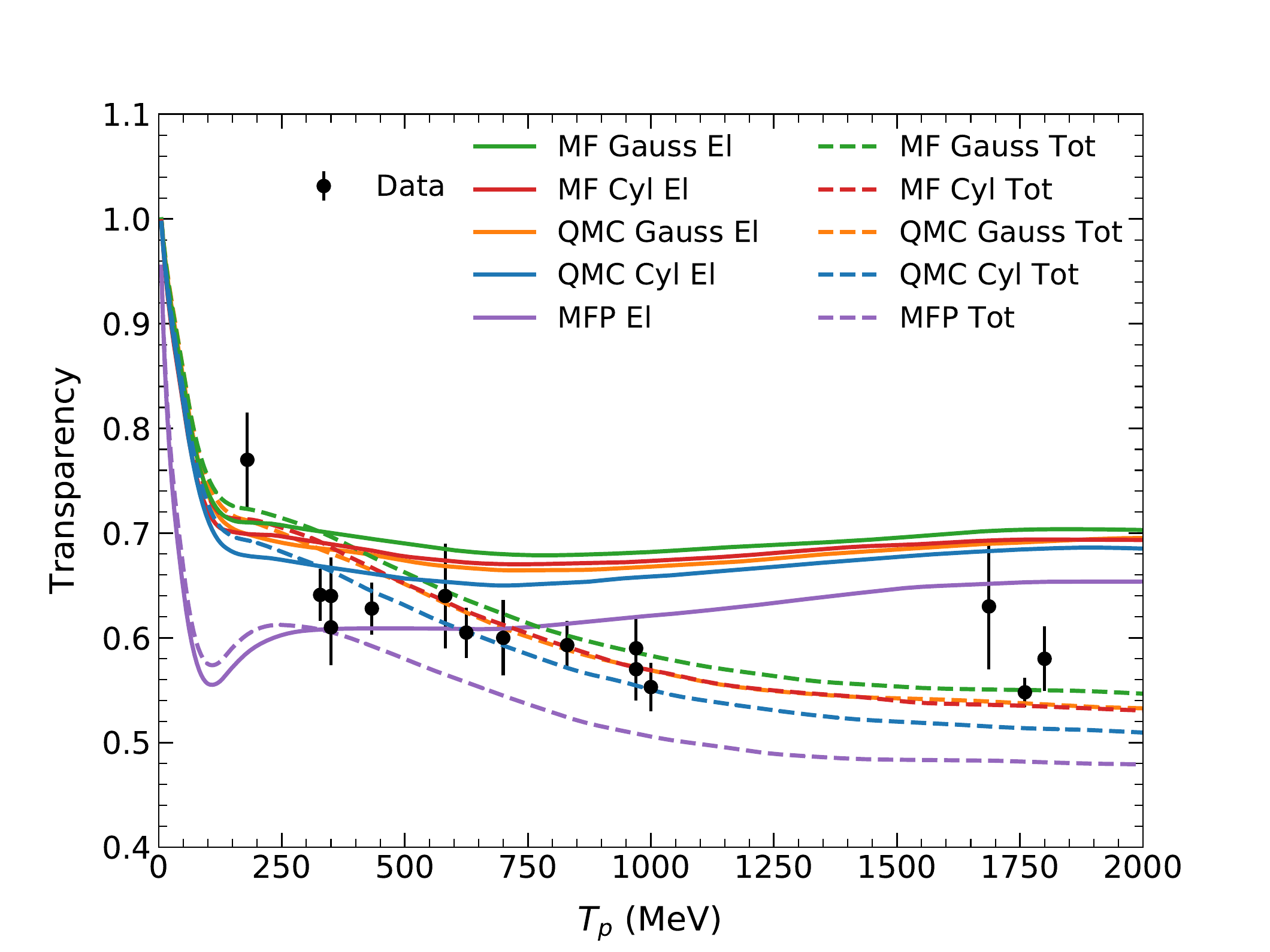}
    \caption{Carbon transparency as a function of the proton kinetic energy. The different curves indicate different approaches used as described in Fig.~\ref{fig:pC_xsec}. The experimental data are taken from Refs.~\cite{ONeill:1994znv, Abbott:1997bc, Rohe:2005vc, Garrow:2001di, Dutta:2003yt, Garino:1992ca} } 
    \label{fig:trans}
\end{figure}

In our Monte Carlo simulation, the nuclear transparency has been obtained via the following procedure.
We randomly sample a nucleon, its position being generated according to either the QMC or MF distributions. We give the nucleon a kick by assigning it a given kinetic energy $T_p$ and three-momentum ${\bf p}$, and propagate it through the nuclear medium. 
The Monte Carlo transparency is then defined as 
\begin{equation}
    T_{\rm MC}= 1-\frac{N_{\rm hits}}{N_{\rm tot}}
\end{equation}
where Pauli blocking has been implemented in the determination of $N_{\rm hits}$. 
Note that for a given initial and final energy and scattering angle of the electron, one can unambiguously define the momentum ${\bf q}$ transferred to the target nucleus. The direction and the momentum of the nucleon in the final state has to be determined applying  energy- and momentum-conservation relations and 
accounting for the Fermi motion of the struck nucleon in the initial state. It follows that defining the kinematics of the hadronic final state after the hard scattering depends on the nuclear model of choice. However, in the analysis of different experiments, the data are given as a function of the average nucleon momentum  (and kinetic energy) given by ${\bf p}={\bf q}$ ($T_p=\sqrt{|{\bf q}|^2 +m_N^2}-m_N$).

In Fig.~\ref{fig:trans} we compare the nuclear transparency data from Refs.~\cite{ONeill:1994znv, Abbott:1997bc} to our predictions.
The different lines are the same as for Fig.~\ref{fig:pC_xsec}. We find an overall satisfactory agreement between the Gaussian and cylinder curves with the experimental data once inelastic effects are taken into consideration;
this corresponds to the results using the NASA parametrization for the nucleon-nucleon cross sections.
For moderate to large values of the proton kinetic energy, pions play an important role in quenching the transparency.
Moreover, the Gaussian and cylinder curves exhibit correct behavior consistent with the data also for small $T_p$ where the simplified MFP model described above fails.
As in Fig.~\ref{fig:pC_xsec}, we observe very small differences between the QMC and MF calculations.
For low and intermediate kinetic energies, the transparency obtained from the MFP approach is much smaller than the corresponding results for the cylinder and Gaussian curves.

\begin{figure}[t]
\includegraphics[width=.23\textwidth]{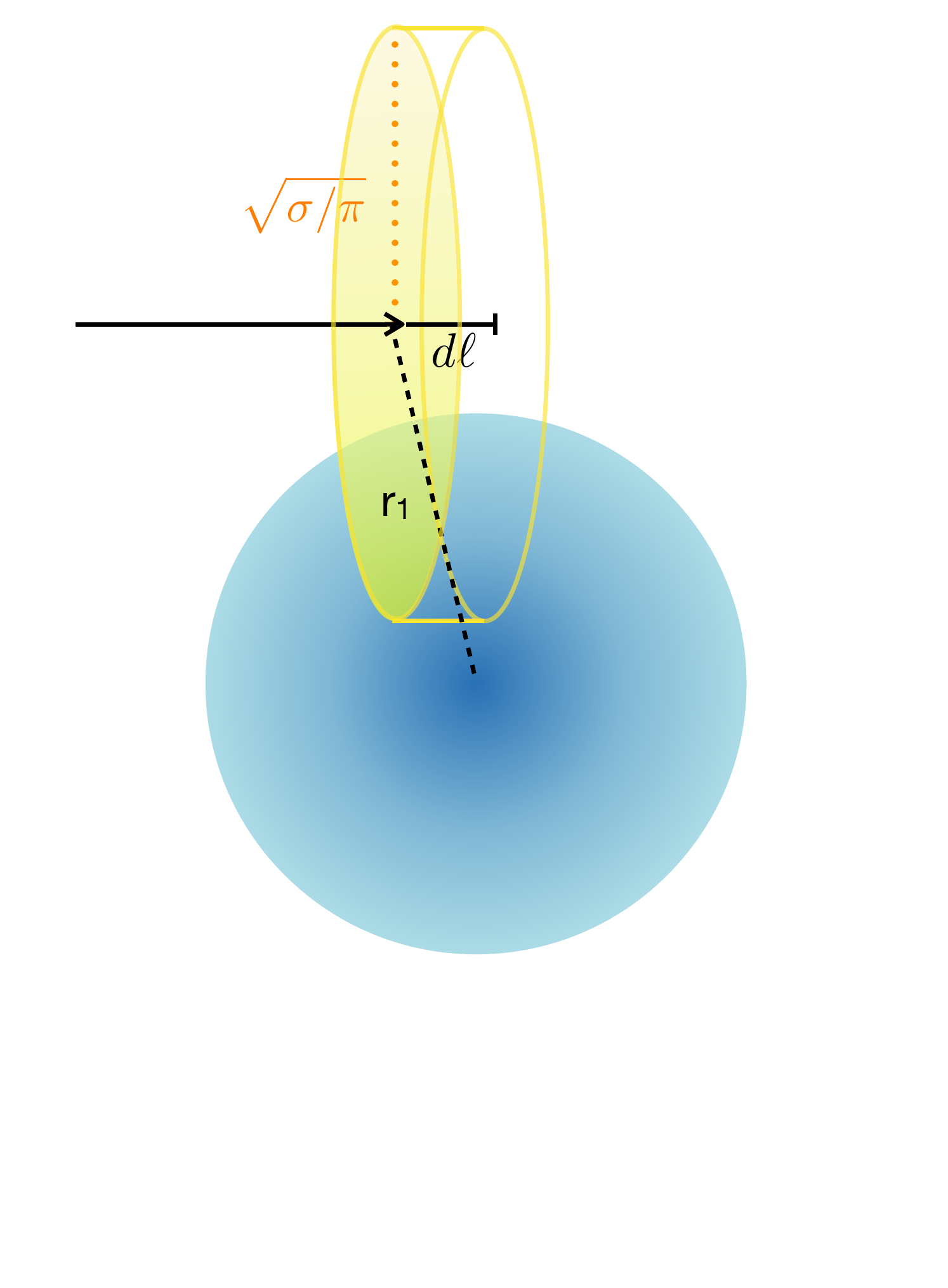}
\includegraphics[width=.21\textwidth]{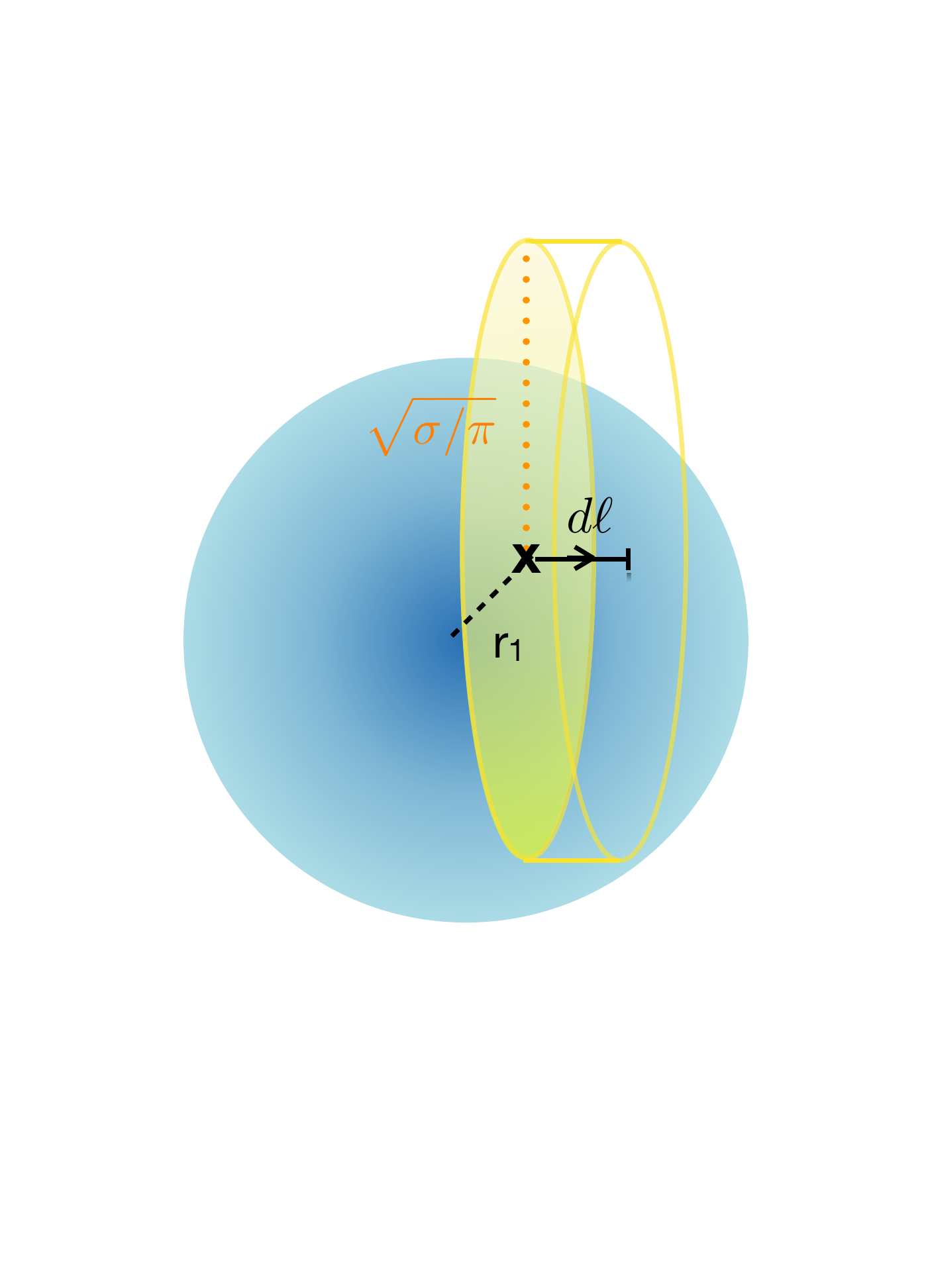}
\caption{Left panel: a schematic picture of an external proton scattering off the nucleus.
The distance from the proton to the center of the nucleus is $r_1$, and the propagation step is $d\ell$.
The radius of the cylinder is given by $\sqrt{\sigma/\pi}$ where $\sigma$ is the interaction cross section between the proton and a background particle; $d\ell$ is also the height of the cylinder.
Right panel: same as for the left one, but for a nucleon kicked inside the nucleus.
This follows what is done in the nuclear transparency event simulations.
\label{fig:cylinder_schematic}}
\end{figure}

Finally, we discuss the origin of the discrepancies between the MFP and the cylinder algorithm with MF configurations for the $p$-carbon cross section and carbon transparency. Both approaches rely on the single-nucleon density distribution to sample the initial nucleon positions (nuclear correlations are neglected) but use different definitions of the interaction probability. 
The left panel of Fig.~\ref{fig:cylinder_schematic} schematically shows one contribution to the $p$-carbon cross section in which the proton is at a distance $r_1$ larger than the nuclear radius. In the cylinder algorithm, the interaction probability is equal to one if a particle is present in the volume defined by: $V=d\ell\cdot \sigma$. Both $\sigma_{pp}$ and $\sigma_{np}$ have a maximum for low  proton momentum values. Hence, for low momenta, the probability of interaction could be non-vanishing even when the projectile proton is far from the center of the nucleus.\\
On the other hand, within the MFP approach, if the probe is outside the nucleus then the approximation of a constant density $\rho(r_1)=0$ within the volume $V = d\ell\cdot \sigma$ yields a vanishing interaction probability. This different behaviour leads to a lower $p$-carbon cross section using the MFP approach, as observed in Fig.~\ref{fig:pC_xsec}. 
When computing the nuclear transparency we kick a nucleon which is located inside the nucleus as displayed in
the right panel of Fig.~\ref{fig:cylinder_schematic}.
In this case, assuming a constant density is more likely to overestimate the interaction probability, especially for low momenta where the cross section is larger.
This observation is consistent with Fig.~\ref{fig:trans} where the MFP curves predict a larger number of interactions, and therefore a lower nuclear transparency, for small $T_p$. 

\subsection{Correlation effects}
\label{sec:corr:eff}

\begin{figure*}[ht]
\includegraphics[width=.45\textwidth]{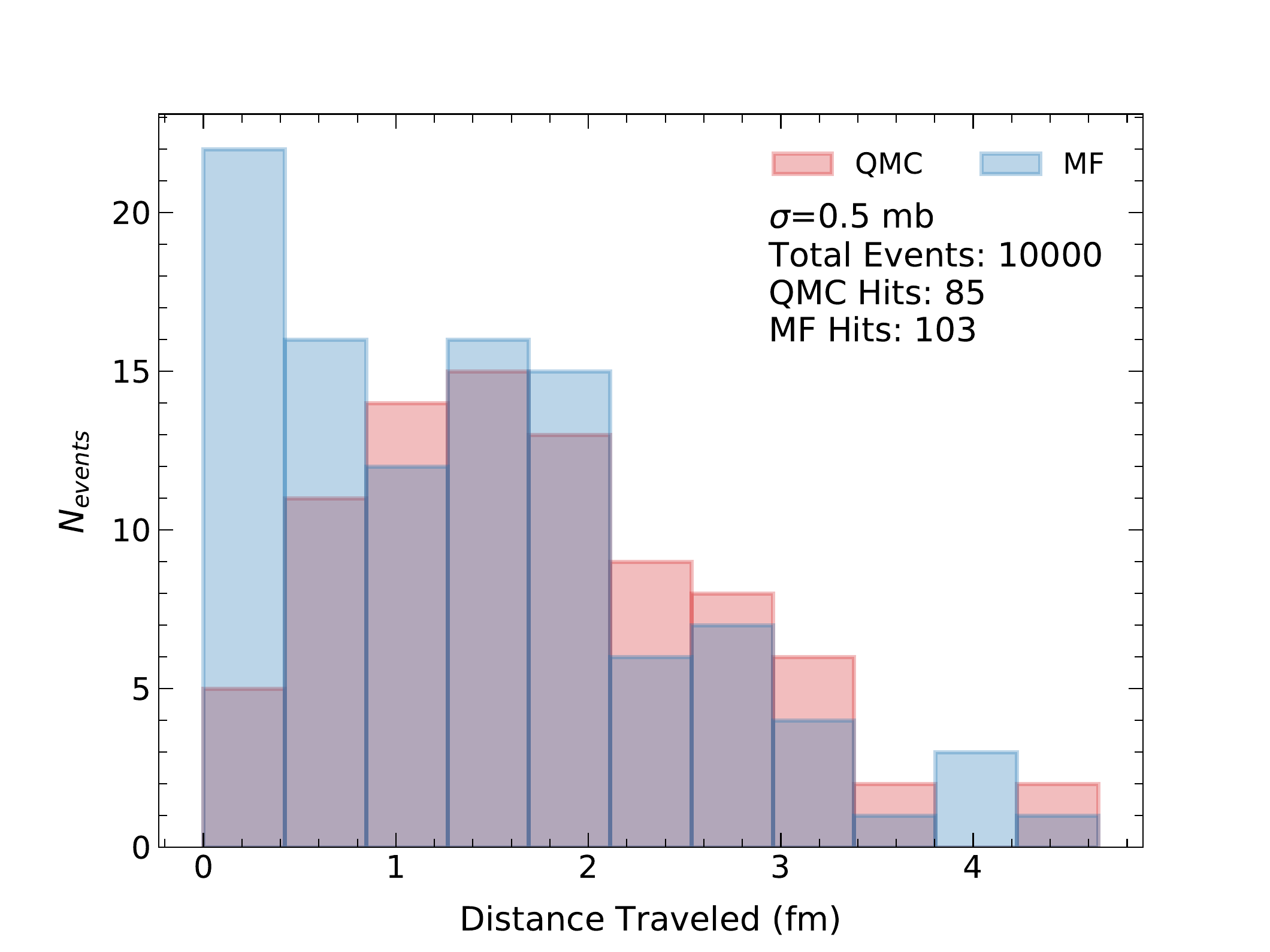}
\includegraphics[width=.45\textwidth]{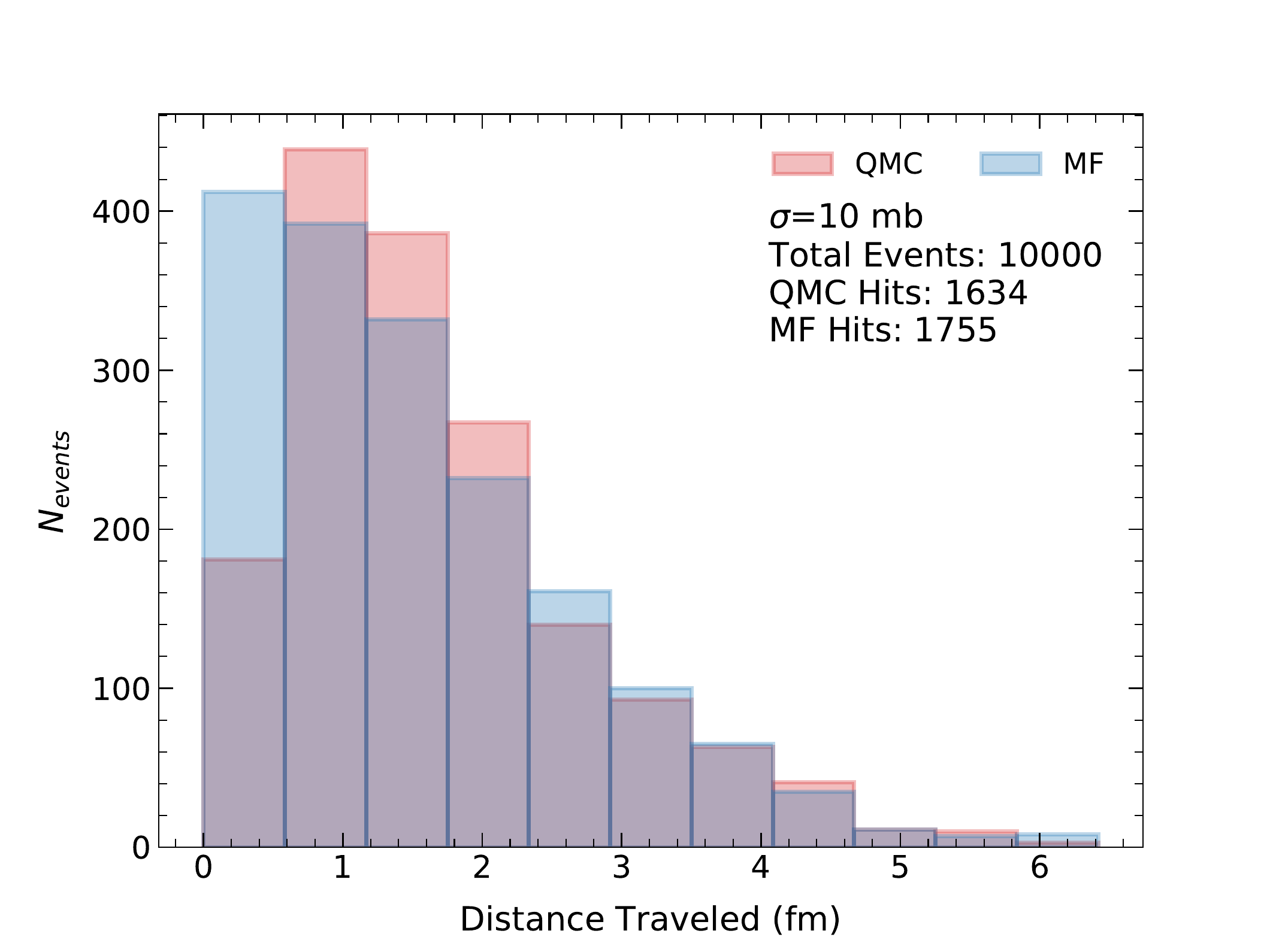}
\\
\includegraphics[width=.45\textwidth]{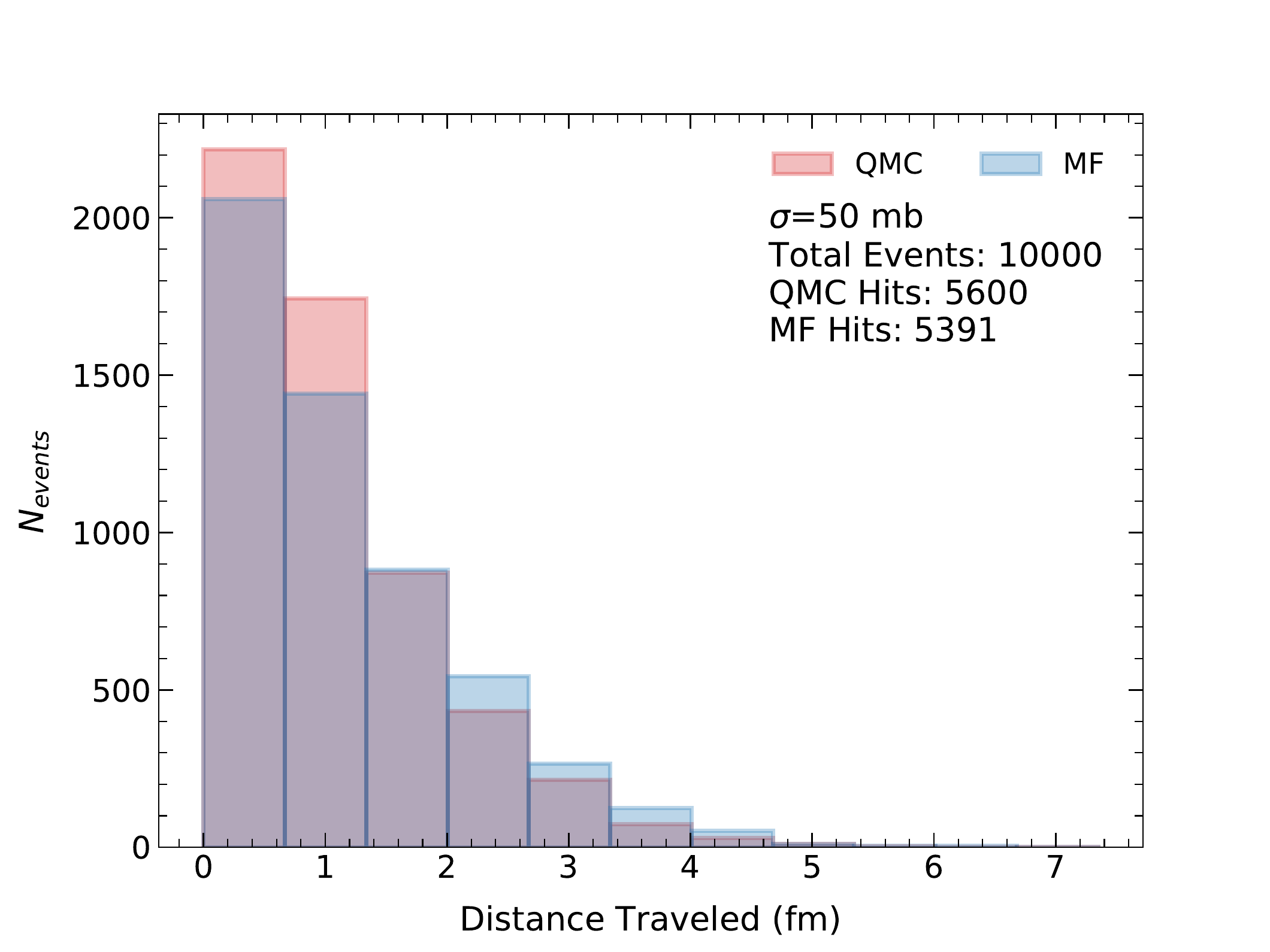}
\includegraphics[width=.45\textwidth]{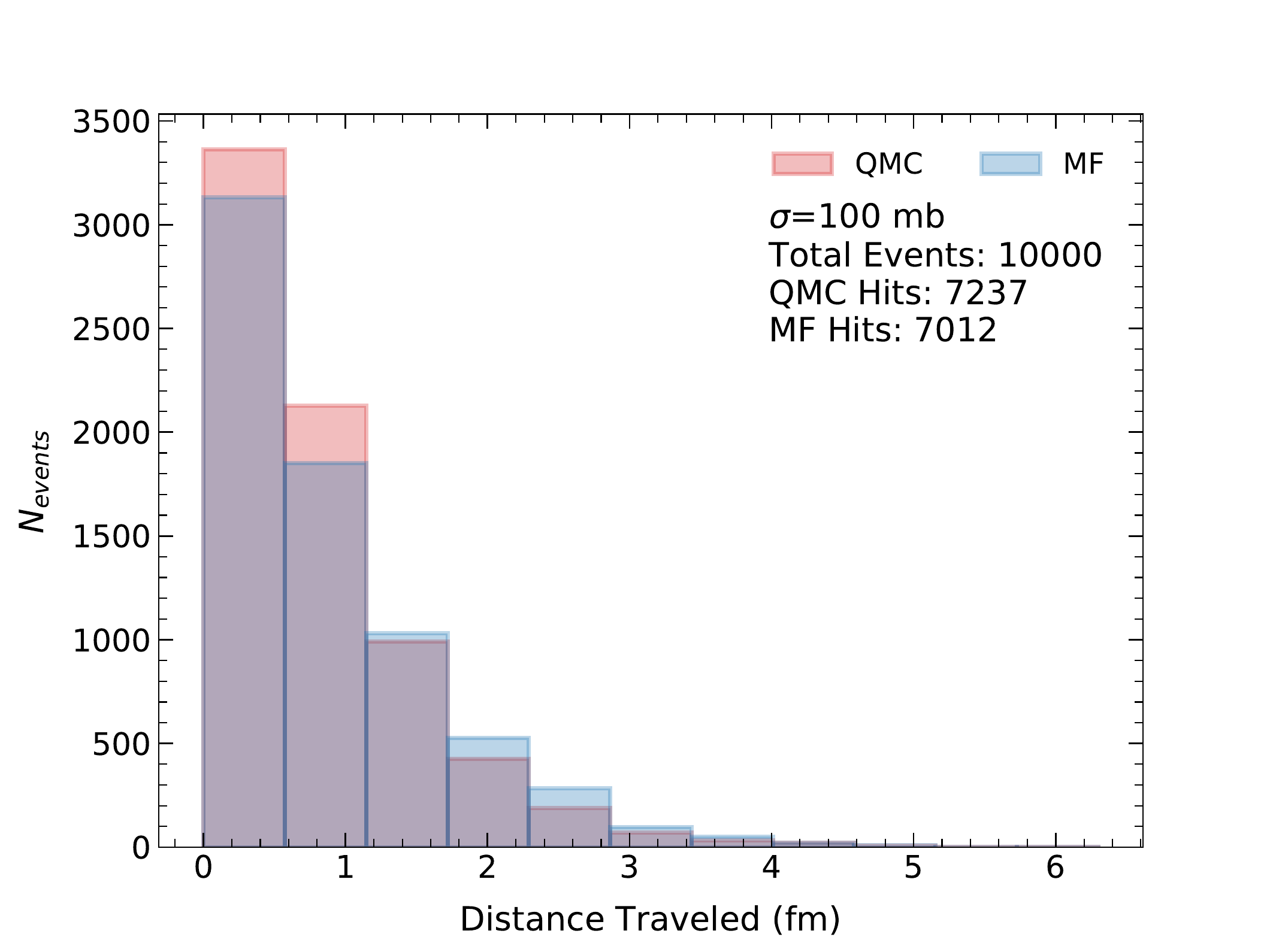}
\caption{The four panels corresponds to histograms of the distance traveled by a struck particle before the first interaction takes place  for different values of the interaction cross section.
The results in blue and red correspond to MF and QMC initial nucleon configurations, respectively.
For each of the panels we also report the fixed cross-section used, the total number of events generated, and the number of hits for each configuration.}
\label{fig:dist}
\end{figure*}

The role played by nuclear correlations in final state interactions of the recoiling nucleon has been investigated in Refs.~\cite{Pandharipande:1992zz,Radici:1994wy,Benhar:1993zz,Nikolaev:1993sj,Benhar:2013dq}.
As discussed in Ref.~\cite{Benhar:1995xa} the hit nucleon is 
surrounded by a short-distance correlation hole produced by both the Pauli principle and the repulsive nature of realistic nuclear interactions. Because of this correlation hole, the stuck nucleon is expected to freely propagate for $\sim 1$ fm before interacting with any of the background particles. To test the validity of these observations in our INC model, in Fig.~\ref{fig:dist} we report the histograms of the distance traveled by a struck nucleon before its first interaction occurs---we stop the simulation afterwards---with each panel corresponding to a different value of the interaction cross section. In order to gauge the effect of nuclear correlations, the initial positions of the nucleons are sampled from either MF (blue) or QMC (red) configurations. A random nucleon inside the nucleus is recoiled and assigned a momentum of 200 MeV. Pauli Blocking has been neglected here to isolate the dependence of the results on the spatial distribution of the nucleons.
We employ the cylinder algorithm and use a fixed cross section---which determines the cylinder base area---varying between 0.5 and 100 mb. 

For $\sigma = 0.5$ and 10 mb, the volume spanned 
by the propagating particle is very small. 
The first and second panels of Fig.~\ref{fig:dist} clearly show the MF distribution peaking toward smaller distances than the QMC distribution.
This difference primarily originates from the short-range repulsion of the AV18 potential that reduces the probability of finding two nucleons close to each other and allows the struck particle to propagate longer before interacting. This effect is more pronounced for cross sections below about $10~{\rm mb}=1~{\rm fm}^2$ since correlations affect nucleon configuration for inter-particle distances within $1\sim2$~fm, as can be seen in Fig.~\ref{fig:density_12b}. On the other hand, larger cross sections yield larger cylinders.  In this case, the propagating particle becomes less sensitive to the local distribution of nucleons and more sensitive to the integrated density in a larger volume, reducing the effect of correlations.
For these larger cross sections, the MF and QMC event distributions follow the same trend, as can be seen in the lower panels of Fig.~\ref{fig:dist}, corresponding to $\sigma=50$ and $100$~mb. 

In each panel we also report the number of hits and the total number of events simulated.
Obviously, the number of interactions increases for the larger cross sections.
It is more interesting to notice that the total number of hits does not present a strong dependence on configurations with (QMC) and without (MF) correlations.
This explains why sampling from these two different set of configurations give comparable results for the $p$-carbon cross section and carbon transparency, displayed in Figs.~\ref{fig:pC_xsec} and \ref{fig:trans}. These observables are indeed more sensitive to the occurrence of any interaction rather than to its location inside the nucleus. This is especially true for the $p$-carbon cross section, since the attenuation of a proton beam is sensitive to the total nuclear density rather than its differential profile (compare, e.g., the solid blue and red lines in Fig.~\ref{fig:pC_xsec}).

An alternative way to include correlation effects in INC models, based on two-body density distributions, has been recently proposed by the NuWro collaboration and leads to a larger enhancement of the nuclear transparency than we find~\cite{Niewczas:2019fro}. This difference can partly be ascribed to correlation terms between two spectator nucleons~\cite{Nikolaev:1993sj} that are automatically incorporated in our QMC configurations, but cannot be modeled by the two-body correlation functions alone. 
In addition, as discussed in Sec.~\ref{sec:corr:eff}, when the nucleon-nucleon cross section is larger than about $20$ mb, the base of the propagating cylinder in our model covers the correlation hole and reduces the importance of correlations. 
This effect is absent when only local properties of two-body distribution functions are considered.

Although nuclear correlations in final state interactions do not play a prominent role in the observables that we have considered, correlations may be important in other experimentally relevant quantities or in high-precision studies.
These observables include proton multiplicity and the distribution of outgoing protons' direction and energy. We plan to continue exploring these questions in future work.

\section{Conclusions}

We have presented a novel INC model that takes as input realistic QMC nuclear configurations to sample the distributions of protons and neutrons inside the nucleus.  Either a cylinder or a Gaussian  distribution is used to define the interaction probability of the propagating nucleon as a function of the impact parameter and the nucleon-nucleon cross sections. We have considered the elastic and the total cross sections, corresponding to the SAID database~\cite{SAID} and to the NASA parametrization~\cite{Norbury:2008}, respectively.

To validate the INC we have compared our simulations with experimental data for $p$-carbon scattering and carbon transparency. For both observables, we find minimal difference between the cylinder and Gaussian distributions, and both correctly reproduce the trend of the data. We gauge the role of nuclear correlations in the propagation of the recoiling nucleons by utilizing nuclear configurations computed within quantum Monte Carlo and comparing with mean field distributions. The distance traveled by a struck particle before the first interaction occurs, for sufficiently small value of the nucleon-nucleon cross section, is significantly longer when nuclear correlations are accounted for. This is due to the short-range repulsion characterizing the AV18 potential that gives rise to a correlation hole surrounding the propagating nucleon, thereby reducing the probability of finding two nucleons close to each other. However, the total number of hits is only mildly affected by the presence of nuclear correlations. As a consequence, they mildly affect our calculations of $p$-carbon scattering and carbon transparency. The results presented for the latter quantity by the NuWro collaboration indicate a larger importance of nuclear correlations, whose implementation is based on the two-body density distributions~\cite{Niewczas:2019fro}.

Nucleon-nucleon cross sections are an important ingredient in INC models. Since the scattering between the two nucleons takes place in the nuclear medium, the free nucleon-nucleon cross sections have to be corrected. In our simulation, we do so by approximately implementing the Pauli exclusion principle: in the aftermath of each collision, the magnitude of final nucleons' momenta have to be larger than the Fermi momentum. This constraint reduces the effective cross section and considerably improves the agreement between our simulations and experimental data. As a next step, along the line of Ref.~\cite{Pandharipande:1992zz}, we plan to replace the bare nucleon mass with an effective one that depends on the momentum and the nuclear density, leading to a more complete and consistent treatment of in-medium effects.  

At energies larger than the pion-production threshold, inelastic contributions to the nucleon-nucleon cross section become relevant. Using the NASA parametrization, which includes these inelasticities, noticeably improves the agreement with $p$-carbon scattering and carbon transparency experimental data in the higher-energy region. Our INC model has yet to include pion propagation in the nuclear medium. Existing event generators have to rely on a number of semi-phenomenological approaches constrained by experimental data~\cite{Salcedo:1987md, Buss:2011mx, Ashery:1981tq,Ashery:1984ne,Jones:1993ps}, the validity of which strongly depends upon the energy of the propagating pion. A consistent implementation of pion propagation and absorption is an extremely challenging problem and we leave this aspect for future works.

The transparency measurements used here depend upon theoretical calculations based on the  plane wave impulse approximation. 
To compare directly to the experimental data, we plan on implementing  the primary interaction vertex using the spectral function formalism~\cite{Benhar:1994hw,Rocco:2015cil,Benhar:2006wy}.
Additionally to further validate our model, we will carry out extensive comparison against available semi-exclusive electron scattering data~\cite{Cruz-Torres:2020uke,Duer:2018sxh}.

\section{Acknowledgments}
We thank John Arrington, Omar Benhar, Donald F. Geesaman, Stefan H\"oche, T.S. Harry Lee, Kajetan Niewczas, Stefan Prestel, Jan Sobczyk, and Mike Wagman for useful discussions   .
PM and AL thank the Argonne Physics Division and the Fermilab theory group, respectively,  for providing consistent office space necessary for the completion of this work. Fermilab is operated by the Fermi Research Alliance, LLC under contract No. DE-AC02-07CH11359 with the United States Department of Energy. The present research is supported by the U.S. Department of Energy, Office of Science, Office of Nuclear Physics, under contracts DE-AC02-06CH11357 (A.L. and N.R.) and by the NUCLEI SciDAC program.

\bibliography{biblio}

\appendix
\section{Nucleon-nucleon interaction Models}

\begin{figure}
    \centering
    \includegraphics[width=0.53\textwidth]{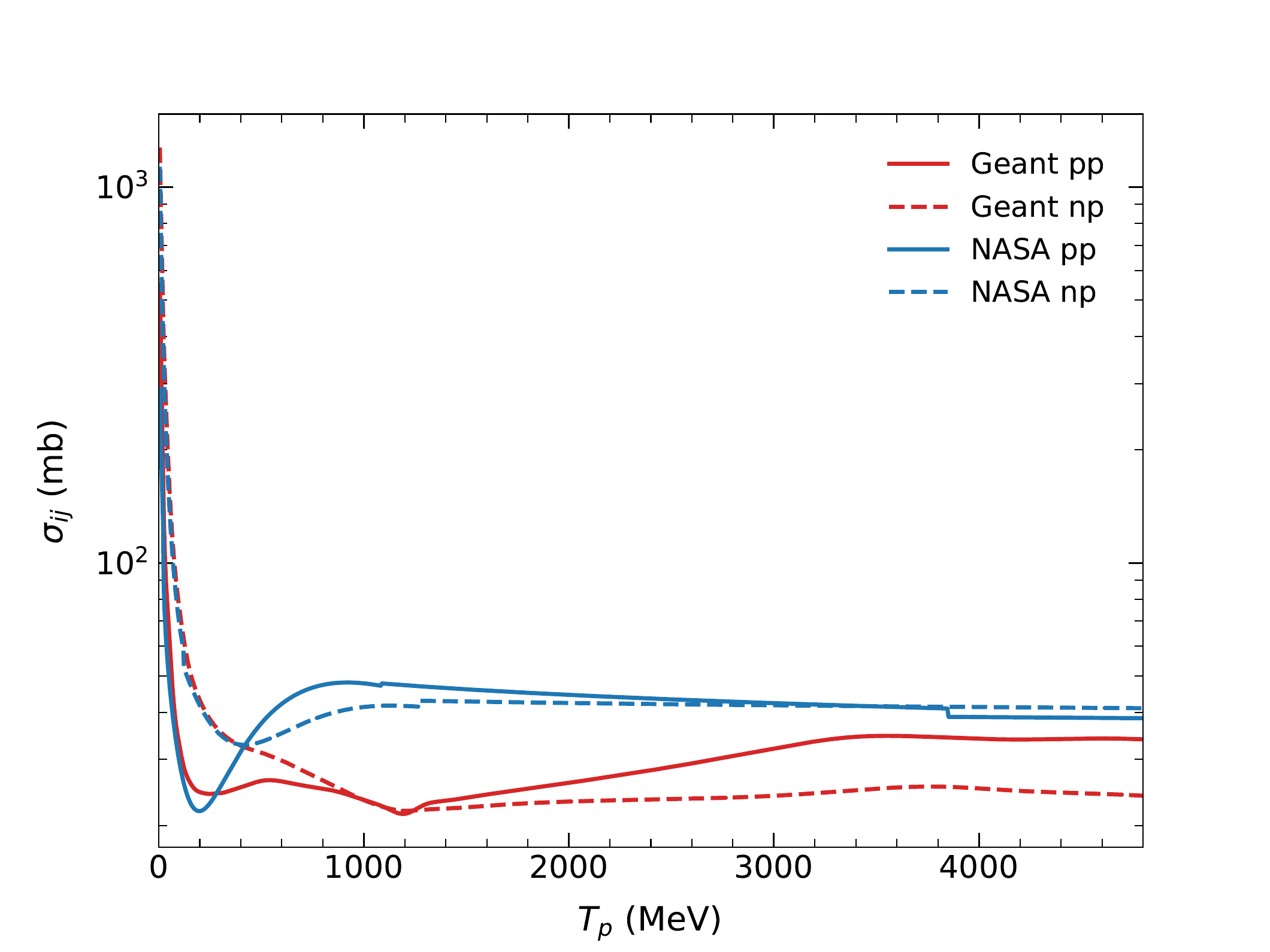}\\
    \includegraphics[width=0.53\textwidth]{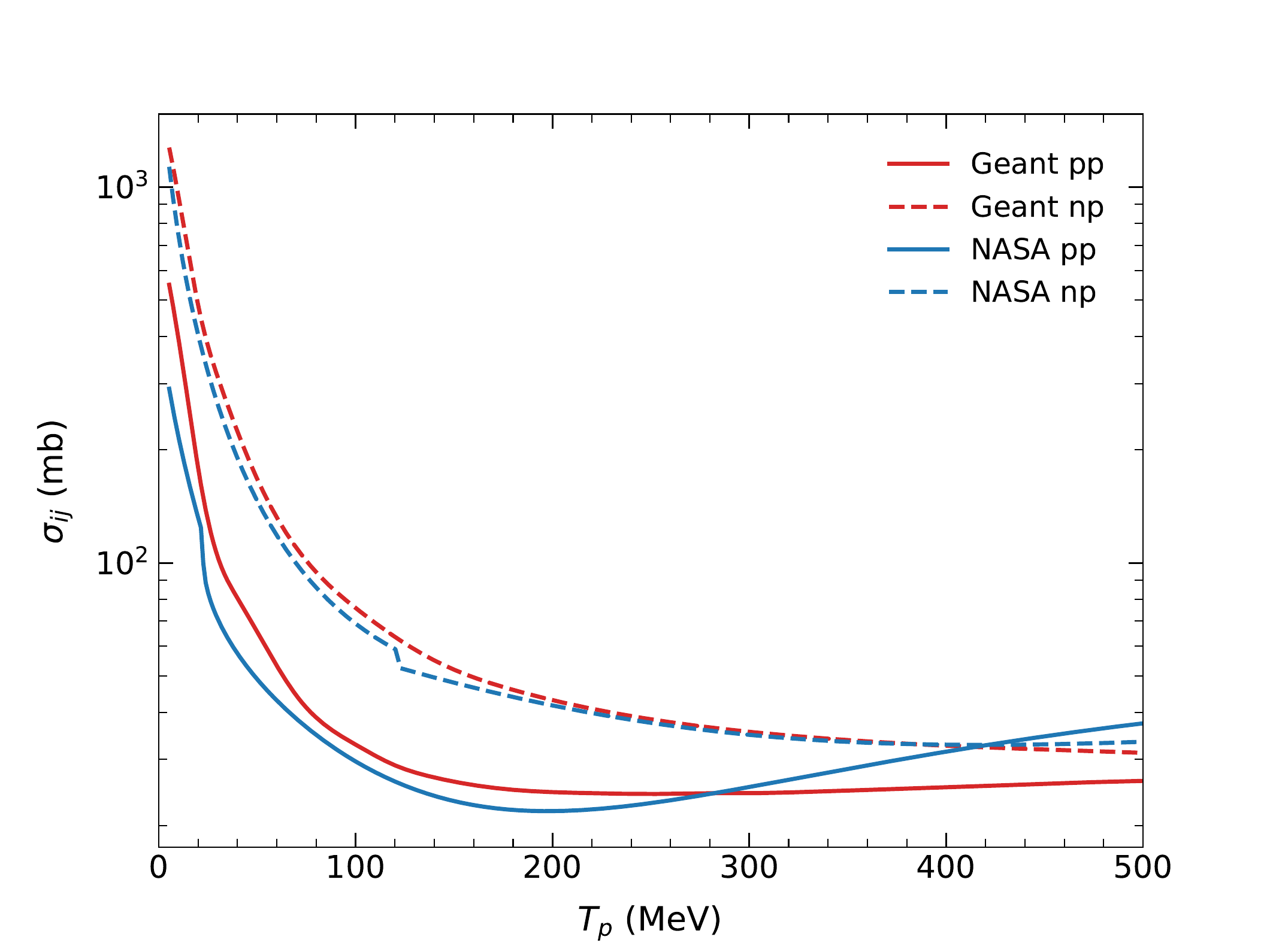}\\
    \caption{Comparison between the two interaction models used within the intranuclear cascade model. The GEANT model is based on the SAID database~\cite{SAID} and obtained using GEANT4~\cite{Agostinelli:2002hh} and the NASA model is given from~\cite{Norbury:2008}.}
    \label{fig:interaction model}
\end{figure}

The nucleon-nucleon cross sections used in our INC are obtained externally.
The elastic cross section is taken from the  SAID database~\cite{SAID}, obtained using \texttt{GEANT4}~\cite{Agostinelli:2002hh}.
The total cross section, including  inelastic contributions (e.g. pion production) is taken from Ref.~\cite{Norbury:2008}.
For reference, we reproduce those in Figs.~\ref{fig:interaction model}.
As it can be seen, the cross sections from these databases are similar in the region where the elastic contribution dominates, that is, below 300~MeV for $pp$ interactions and 500~MeV for $np$ interactions.
The differences at high energy can be attributed to inelastic contributions.

\end{document}

%% file: tikz/Algorithm.tex
\tikzset{
  decision/.style={
        diamond,
        draw, thick,
        text width=5.5em,
        text badly centered,
        inner sep=0pt
    },
    block/.style={
        rectangle,
        draw, thick,
        text width=10em,
        text centered,
        rounded corners
    },
    cloud/.style={
        draw,
        ellipse,
        minimum height=2em
    },
    descr/.style={
        fill=white,
        inner sep=2.5pt
    },
    connector/.style={
        -latex,
        font=\scriptsize
    },
    rectangle connector/.style={
        connector,
        to path={(\tikztostart) -- ++(#1,0pt) \tikztonodes |- (\tikztotarget) },
        pos=0.5
    },
    rectangle connector/.default=-2cm,
    straight connector/.style={
        connector,
        to path=--(\tikztotarget) \tikztonodes
    },
    line/.style={>=latex,->,thick},
    evey edge quotes/.style = {auto=right},
}

\begin{figure*}
    \centering
    \tikzsetnextfilename{algorithm}
    \begin{tikzpicture}
        \matrix (m) [matrix of nodes, column sep=1cm, row sep=6mm, align=center, nodes={anchor=center}, nodes in empty cells]    
        {
            |[block]| {Start} & &\\
            |[block]| {Generate nuclear configuration} & &\\
            |[block]| {Kick random nucleon} & &\\
            |[block]| {Evolve propagating nucleon(s) for time $\delta t$} & &\\
            |[decision]| {Possible interaction?} & |[decision]| {Outside nucleus?} & |[block]| {Remove from propagating particle list}\\
            |[block]| {Generate outgoing momenta} &  & \\
            |[decision]| {Pauli Blocked?} & |[block]|{Restore original outgoing momenta} & & \\
            |[block]| {Update outgoing momenta of particles} & &|[block]| {Add particle to list and update formation zone}& \\
        };
        \path[line] (m-1-1) edge (m-2-1);
        \path[line] (m-2-1) edge (m-3-1);
        \path[line] (m-3-1) edge (m-4-1);
        \path[line] (m-4-1) edge (m-5-1);
        \path[line] (m-5-1) edge node[midway, above, text=black]{No} (m-5-2);
        \path[line] (m-5-1) edge node[midway, right, text=black]{Yes} (m-6-1);
        \path[line] (m-5-2) edge node[midway, above, text=black]{Yes} (m-5-3);
        \draw[line] (m-5-2) |- node[midway, right, text=black]{No} (m-4-1);
        \path[line] (m-6-1) edge (m-7-1);
        \path[line] (m-7-1) edge node[midway, above, text=black]{Yes} (m-7-2);
        \path[line] (m-7-2) edge (m-5-2);
        \path[line] (m-7-1) edge node[midway, right, text=black]{No} (m-8-1);
        \path[line] (m-8-1) edge (m-8-3);
        \draw[line] (m-8-3) -- (m-7-3.center) -- (m-6-3.center) -- (m-6-2.center) -- (m-5-2);
    \end{tikzpicture}
    \caption{
        The proposed algorithm for the INC model.
        \label{fig:algorithm}
    }
\end{figure*}

%% file: main.bbl
\begin{thebibliography}{90}%
\makeatletter
\providecommand \@ifxundefined [1]{%
 \@ifx{#1\undefined}
}%
\providecommand \@ifnum [1]{%
 \ifnum #1\expandafter \@firstoftwo
 \else \expandafter \@secondoftwo
 \fi
}%
\providecommand \@ifx [1]{%
 \ifx #1\expandafter \@firstoftwo
 \else \expandafter \@secondoftwo
 \fi
}%
\providecommand \natexlab [1]{#1}%
\providecommand \enquote  [1]{``#1''}%
\providecommand \bibnamefont  [1]{#1}%
\providecommand \bibfnamefont [1]{#1}%
\providecommand \citenamefont [1]{#1}%
\providecommand \href@noop [0]{\@secondoftwo}%
\providecommand \href [0]{\begingroup \@sanitize@url \@href}%
\providecommand \@href[1]{\@@startlink{#1}\@@href}%
\providecommand \@@href[1]{\endgroup#1\@@endlink}%
\providecommand \@sanitize@url [0]{\catcode `\\12\catcode `\$12\catcode
  `\&12\catcode `\#12\catcode `\^12\catcode `\_12\catcode `\%12\relax}%
\providecommand \@@startlink[1]{}%
\providecommand \@@endlink[0]{}%
\providecommand \url  [0]{\begingroup\@sanitize@url \@url }%
\providecommand \@url [1]{\endgroup\@href {#1}{\urlprefix }}%
\providecommand \urlprefix  [0]{URL }%
\providecommand \Eprint [0]{\href }%
\providecommand \doibase [0]{http://dx.doi.org/}%
\providecommand \selectlanguage [0]{\@gobble}%
\providecommand \bibinfo  [0]{\@secondoftwo}%
\providecommand \bibfield  [0]{\@secondoftwo}%
\providecommand \translation [1]{[#1]}%
\providecommand \BibitemOpen [0]{}%
\providecommand \bibitemStop [0]{}%
\providecommand \bibitemNoStop [0]{.\EOS\space}%
\providecommand \EOS [0]{\spacefactor3000\relax}%
\providecommand \BibitemShut  [1]{\csname bibitem#1\endcsname}%
\let\auto@bib@innerbib\@empty
\bibitem [{\citenamefont {Barreau}\ \emph {et~al.}(1983)\citenamefont {Barreau}
  \emph {et~al.}}]{Barreau:1983ht}%
  \BibitemOpen
  \bibfield  {author} {\bibinfo {author} {\bibfnamefont {P.}~\bibnamefont
  {Barreau}} \emph {et~al.},\ }\href {\doibase 10.1016/0375-9474(83)90217-8}
  {\bibfield  {journal} {\bibinfo  {journal} {Nucl. Phys. A}\ }\textbf
  {\bibinfo {volume} {402}},\ \bibinfo {pages} {515} (\bibinfo {year}
  {1983})}\BibitemShut {NoStop}%
\bibitem [{\citenamefont {Makins}\ \emph {et~al.}(1994)\citenamefont {Makins}
  \emph {et~al.}}]{Makins:1994mm}%
  \BibitemOpen
  \bibfield  {author} {\bibinfo {author} {\bibfnamefont {N.}~\bibnamefont
  {Makins}} \emph {et~al.},\ }\href {\doibase 10.1103/PhysRevLett.72.1986}
  {\bibfield  {journal} {\bibinfo  {journal} {Phys. Rev. Lett.}\ }\textbf
  {\bibinfo {volume} {72}},\ \bibinfo {pages} {1986} (\bibinfo {year}
  {1994})}\BibitemShut {NoStop}%
\bibitem [{\citenamefont {Anghinolfi}\ \emph {et~al.}(1995)\citenamefont
  {Anghinolfi} \emph {et~al.}}]{Anghinolfi:1995bz}%
  \BibitemOpen
  \bibfield  {author} {\bibinfo {author} {\bibfnamefont {M.}~\bibnamefont
  {Anghinolfi}} \emph {et~al.},\ }\href {\doibase 10.1088/0954-3899/21/3/001}
  {\bibfield  {journal} {\bibinfo  {journal} {J. Phys. G}\ }\textbf {\bibinfo
  {volume} {21}},\ \bibinfo {pages} {L9} (\bibinfo {year} {1995})}\BibitemShut
  {NoStop}%
\bibitem [{\citenamefont {Abbott}\ \emph {et~al.}(1998)\citenamefont {Abbott}
  \emph {et~al.}}]{Abbott:1997bc}%
  \BibitemOpen
  \bibfield  {author} {\bibinfo {author} {\bibfnamefont {D.}~\bibnamefont
  {Abbott}} \emph {et~al.},\ }\href {\doibase 10.1103/PhysRevLett.80.5072}
  {\bibfield  {journal} {\bibinfo  {journal} {Phys. Rev. Lett.}\ }\textbf
  {\bibinfo {volume} {80}},\ \bibinfo {pages} {5072} (\bibinfo {year}
  {1998})}\BibitemShut {NoStop}%
\bibitem [{\citenamefont {Dutta}\ \emph {et~al.}(2000)\citenamefont {Dutta}
  \emph {et~al.}}]{Dutta:2000sn}%
  \BibitemOpen
  \bibfield  {author} {\bibinfo {author} {\bibfnamefont {D.}~\bibnamefont
  {Dutta}} \emph {et~al.},\ }\href {\doibase 10.1103/PhysRevC.61.061602}
  {\bibfield  {journal} {\bibinfo  {journal} {Phys. Rev. C}\ }\textbf {\bibinfo
  {volume} {61}},\ \bibinfo {pages} {061602} (\bibinfo {year}
  {2000})}\BibitemShut {NoStop}%
\bibitem [{\citenamefont {Dutta}\ \emph {et~al.}(2003)\citenamefont {Dutta}
  \emph {et~al.}}]{Dutta:2003yt}%
  \BibitemOpen
  \bibfield  {author} {\bibinfo {author} {\bibfnamefont {D.}~\bibnamefont
  {Dutta}} \emph {et~al.} (\bibinfo {collaboration} {JLab E91013}),\ }\href
  {\doibase 10.1103/PhysRevC.68.064603} {\bibfield  {journal} {\bibinfo
  {journal} {Phys. Rev. C}\ }\textbf {\bibinfo {volume} {68}},\ \bibinfo
  {pages} {064603} (\bibinfo {year} {2003})},\ \Eprint
  {http://arxiv.org/abs/nucl-ex/0303011} {arXiv:nucl-ex/0303011} \BibitemShut
  {NoStop}%
\bibitem [{\citenamefont {Rohe}\ \emph {et~al.}(2005)\citenamefont {Rohe} \emph
  {et~al.}}]{Rohe:2005vc}%
  \BibitemOpen
  \bibfield  {author} {\bibinfo {author} {\bibfnamefont {D.}~\bibnamefont
  {Rohe}} \emph {et~al.} (\bibinfo {collaboration} {E97-006}),\ }\href
  {\doibase 10.1103/PhysRevC.72.054602} {\bibfield  {journal} {\bibinfo
  {journal} {Phys. Rev. C}\ }\textbf {\bibinfo {volume} {72}},\ \bibinfo
  {pages} {054602} (\bibinfo {year} {2005})},\ \Eprint
  {http://arxiv.org/abs/nucl-ex/0506007} {arXiv:nucl-ex/0506007} \BibitemShut
  {NoStop}%
\bibitem [{\citenamefont {Dai}\ \emph {et~al.}(2018)\citenamefont {Dai} \emph
  {et~al.}}]{Dai:2018xhi}%
  \BibitemOpen
  \bibfield  {author} {\bibinfo {author} {\bibfnamefont {H.}~\bibnamefont
  {Dai}} \emph {et~al.} (\bibinfo {collaboration} {Jefferson Lab Hall A}),\
  }\href {\doibase 10.1103/PhysRevC.98.014617} {\bibfield  {journal} {\bibinfo
  {journal} {Phys. Rev. C}\ }\textbf {\bibinfo {volume} {98}},\ \bibinfo
  {pages} {014617} (\bibinfo {year} {2018})},\ \Eprint
  {http://arxiv.org/abs/1803.01910} {arXiv:1803.01910 [nucl-ex]} \BibitemShut
  {NoStop}%
\bibitem [{\citenamefont {Dai}\ \emph {et~al.}(2019)\citenamefont {Dai} \emph
  {et~al.}}]{Dai:2018gch}%
  \BibitemOpen
  \bibfield  {author} {\bibinfo {author} {\bibfnamefont {H.}~\bibnamefont
  {Dai}} \emph {et~al.},\ }\href {\doibase 10.1103/PhysRevC.99.054608}
  {\bibfield  {journal} {\bibinfo  {journal} {Phys. Rev. C}\ }\textbf {\bibinfo
  {volume} {99}},\ \bibinfo {pages} {054608} (\bibinfo {year} {2019})},\
  \Eprint {http://arxiv.org/abs/1810.10575} {arXiv:1810.10575 [nucl-ex]}
  \BibitemShut {NoStop}%
\bibitem [{\citenamefont {Murphy}\ \emph {et~al.}(2019)\citenamefont {Murphy}
  \emph {et~al.}}]{Murphy:2019wed}%
  \BibitemOpen
  \bibfield  {author} {\bibinfo {author} {\bibfnamefont {M.}~\bibnamefont
  {Murphy}} \emph {et~al.},\ }\href {\doibase 10.1103/PhysRevC.100.054606}
  {\bibfield  {journal} {\bibinfo  {journal} {Phys. Rev. C}\ }\textbf {\bibinfo
  {volume} {100}},\ \bibinfo {pages} {054606} (\bibinfo {year} {2019})},\
  \Eprint {http://arxiv.org/abs/1908.01802} {arXiv:1908.01802 [hep-ex]}
  \BibitemShut {NoStop}%
\bibitem [{\citenamefont {Abe}\ \emph {et~al.}(2016)\citenamefont {Abe} \emph
  {et~al.}}]{Abe:2016tmq}%
  \BibitemOpen
  \bibfield  {author} {\bibinfo {author} {\bibfnamefont {K.}~\bibnamefont
  {Abe}} \emph {et~al.} (\bibinfo {collaboration} {T2K}),\ }\href {\doibase
  10.1103/PhysRevD.93.112012} {\bibfield  {journal} {\bibinfo  {journal} {Phys.
  Rev.}\ }\textbf {\bibinfo {volume} {D93}},\ \bibinfo {pages} {112012}
  (\bibinfo {year} {2016})},\ \Eprint {http://arxiv.org/abs/1602.03652}
  {arXiv:1602.03652 [hep-ex]} \BibitemShut {NoStop}%
\bibitem [{\citenamefont {Ayres}\ \emph {et~al.}(2007)\citenamefont {Ayres}
  \emph {et~al.}}]{Ayres:2007tu}%
  \BibitemOpen
  \bibfield  {author} {\bibinfo {author} {\bibfnamefont {D.~S.}\ \bibnamefont
  {Ayres}} \emph {et~al.} (\bibinfo {collaboration} {NOvA}),\ }\href {\doibase
  10.2172/935497} {\  (\bibinfo {year} {2007}),\ 10.2172/935497}\BibitemShut
  {NoStop}%
\bibitem [{\citenamefont {Abe}\ \emph {et~al.}(2018)\citenamefont {Abe} \emph
  {et~al.}}]{Abe:2018uyc}%
  \BibitemOpen
  \bibfield  {author} {\bibinfo {author} {\bibfnamefont {K.}~\bibnamefont
  {Abe}} \emph {et~al.} (\bibinfo {collaboration} {Hyper-Kamiokande}),\
  }\href@noop {} {\  (\bibinfo {year} {2018})},\ \Eprint
  {http://arxiv.org/abs/1805.04163} {arXiv:1805.04163 [physics.ins-det]}
  \BibitemShut {NoStop}%
\bibitem [{\citenamefont {Abi}\ \emph {et~al.}(2020)\citenamefont {Abi} \emph
  {et~al.}}]{Abi:2020wmh}%
  \BibitemOpen
  \bibfield  {author} {\bibinfo {author} {\bibfnamefont {B.}~\bibnamefont
  {Abi}} \emph {et~al.} (\bibinfo {collaboration} {DUNE}),\ }\href@noop {} {\
  (\bibinfo {year} {2020})},\ \Eprint {http://arxiv.org/abs/2002.02967}
  {arXiv:2002.02967 [physics.ins-det]} \BibitemShut {NoStop}%
\bibitem [{\citenamefont {Åkesson}\ \emph {et~al.}(2018)\citenamefont
  {Åkesson} \emph {et~al.}}]{Akesson:2018vlm}%
  \BibitemOpen
  \bibfield  {author} {\bibinfo {author} {\bibfnamefont {T.}~\bibnamefont
  {Åkesson}} \emph {et~al.} (\bibinfo {collaboration} {LDMX}),\ }\href@noop {}
  {\  (\bibinfo {year} {2018})},\ \Eprint {http://arxiv.org/abs/1808.05219}
  {arXiv:1808.05219 [hep-ex]} \BibitemShut {NoStop}%
\bibitem [{\citenamefont {Palamara}(2016)}]{Palamara:2016uqu}%
  \BibitemOpen
  \bibfield  {author} {\bibinfo {author} {\bibfnamefont {O.}~\bibnamefont
  {Palamara}} (\bibinfo {collaboration} {ArgoNeuT}),\ }\bibfield  {booktitle}
  {\emph {\bibinfo {booktitle} {{Proceedings, 10th International Workshop on
  Neutrino-Nucleus Interactions in the Few Region (NuInt15): Osaka, Japan,
  November 16-21, 2015}}},\ }\href {\doibase 10.7566/JPSCP.12.010017}
  {\bibfield  {journal} {\bibinfo  {journal} {JPS Conf. Proc.}\ }\textbf
  {\bibinfo {volume} {12}},\ \bibinfo {pages} {010017} (\bibinfo {year}
  {2016})}\BibitemShut {NoStop}%
\bibitem [{\citenamefont {Kelly}\ \emph {et~al.}(2019)\citenamefont {Kelly},
  \citenamefont {Machado}, \citenamefont {Martinez~Soler}, \citenamefont
  {Parke},\ and\ \citenamefont {Perez~Gonzalez}}]{Kelly:2019itm}%
  \BibitemOpen
  \bibfield  {author} {\bibinfo {author} {\bibfnamefont {K.~J.}\ \bibnamefont
  {Kelly}}, \bibinfo {author} {\bibfnamefont {P.~A.}\ \bibnamefont {Machado}},
  \bibinfo {author} {\bibfnamefont {I.}~\bibnamefont {Martinez~Soler}},
  \bibinfo {author} {\bibfnamefont {S.~J.}\ \bibnamefont {Parke}}, \ and\
  \bibinfo {author} {\bibfnamefont {Y.~F.}\ \bibnamefont {Perez~Gonzalez}},\
  }\href {\doibase 10.1103/PhysRevLett.123.081801} {\bibfield  {journal}
  {\bibinfo  {journal} {Phys. Rev. Lett.}\ }\textbf {\bibinfo {volume} {123}},\
  \bibinfo {pages} {081801} (\bibinfo {year} {2019})},\ \Eprint
  {http://arxiv.org/abs/1904.02751} {arXiv:1904.02751 [hep-ph]} \BibitemShut
  {NoStop}%
\bibitem [{\citenamefont {Rott}\ \emph {et~al.}(2017)\citenamefont {Rott},
  \citenamefont {In}, \citenamefont {Kumar},\ and\ \citenamefont
  {Yaylali}}]{Rott:2016mzs}%
  \BibitemOpen
  \bibfield  {author} {\bibinfo {author} {\bibfnamefont {C.}~\bibnamefont
  {Rott}}, \bibinfo {author} {\bibfnamefont {S.}~\bibnamefont {In}}, \bibinfo
  {author} {\bibfnamefont {J.}~\bibnamefont {Kumar}}, \ and\ \bibinfo {author}
  {\bibfnamefont {D.}~\bibnamefont {Yaylali}},\ }\href {\doibase
  10.1088/1475-7516/2017/01/016} {\bibfield  {journal} {\bibinfo  {journal}
  {JCAP}\ }\textbf {\bibinfo {volume} {01}},\ \bibinfo {pages} {016} (\bibinfo
  {year} {2017})},\ \Eprint {http://arxiv.org/abs/1609.04876} {arXiv:1609.04876
  [hep-ph]} \BibitemShut {NoStop}%
\bibitem [{\citenamefont {Rott}\ \emph {et~al.}(2019)\citenamefont {Rott},
  \citenamefont {Jeong}, \citenamefont {Kumar},\ and\ \citenamefont
  {Yaylali}}]{Rott:2019stu}%
  \BibitemOpen
  \bibfield  {author} {\bibinfo {author} {\bibfnamefont {C.}~\bibnamefont
  {Rott}}, \bibinfo {author} {\bibfnamefont {D.}~\bibnamefont {Jeong}},
  \bibinfo {author} {\bibfnamefont {J.}~\bibnamefont {Kumar}}, \ and\ \bibinfo
  {author} {\bibfnamefont {D.}~\bibnamefont {Yaylali}},\ }\href {\doibase
  10.1088/1475-7516/2019/07/006} {\bibfield  {journal} {\bibinfo  {journal}
  {JCAP}\ }\textbf {\bibinfo {volume} {07}},\ \bibinfo {pages} {006} (\bibinfo
  {year} {2019})},\ \Eprint {http://arxiv.org/abs/1903.04175} {arXiv:1903.04175
  [astro-ph.HE]} \BibitemShut {NoStop}%
\bibitem [{\citenamefont {Serber}(1947)}]{Serber:1947zza}%
  \BibitemOpen
  \bibfield  {author} {\bibinfo {author} {\bibfnamefont {R.}~\bibnamefont
  {Serber}},\ }\href {\doibase 10.1103/PhysRev.72.1114} {\bibfield  {journal}
  {\bibinfo  {journal} {Phys. Rev.}\ }\textbf {\bibinfo {volume} {72}},\
  \bibinfo {pages} {1114} (\bibinfo {year} {1947})}\BibitemShut {NoStop}%
\bibitem [{\citenamefont {Metropolis}\ \emph {et~al.}(1958)\citenamefont
  {Metropolis}, \citenamefont {Bivins}, \citenamefont {Storm}, \citenamefont
  {Miller}, \citenamefont {Friedlander},\ and\ \citenamefont
  {Turkevich}}]{Metropolis:1958sb}%
  \BibitemOpen
  \bibfield  {author} {\bibinfo {author} {\bibfnamefont {N.}~\bibnamefont
  {Metropolis}}, \bibinfo {author} {\bibfnamefont {R.}~\bibnamefont {Bivins}},
  \bibinfo {author} {\bibfnamefont {M.}~\bibnamefont {Storm}}, \bibinfo
  {author} {\bibfnamefont {J.}~\bibnamefont {Miller}}, \bibinfo {author}
  {\bibfnamefont {G.}~\bibnamefont {Friedlander}}, \ and\ \bibinfo {author}
  {\bibfnamefont {A.}~\bibnamefont {Turkevich}},\ }\href {\doibase
  10.1103/PhysRev.110.204} {\bibfield  {journal} {\bibinfo  {journal} {Phys.
  Rev.}\ }\textbf {\bibinfo {volume} {110}},\ \bibinfo {pages} {204} (\bibinfo
  {year} {1958})}\BibitemShut {NoStop}%
\bibitem [{\citenamefont {Bertini}(1963)}]{Bertini:1963zzc}%
  \BibitemOpen
  \bibfield  {author} {\bibinfo {author} {\bibfnamefont {H.~W.}\ \bibnamefont
  {Bertini}},\ }\href {\doibase 10.1103/PhysRev.131.1801} {\bibfield  {journal}
  {\bibinfo  {journal} {Phys. Rev.}\ }\textbf {\bibinfo {volume} {131}},\
  \bibinfo {pages} {1801} (\bibinfo {year} {1963})}\BibitemShut {NoStop}%
\bibitem [{\citenamefont {Cugnon}(1980)}]{Cugnon:1980zz}%
  \BibitemOpen
  \bibfield  {author} {\bibinfo {author} {\bibfnamefont {J.}~\bibnamefont
  {Cugnon}},\ }\href {\doibase 10.1103/PhysRevC.22.1885} {\bibfield  {journal}
  {\bibinfo  {journal} {Phys. Rev. C}\ }\textbf {\bibinfo {volume} {22}},\
  \bibinfo {pages} {1885} (\bibinfo {year} {1980})}\BibitemShut {NoStop}%
\bibitem [{\citenamefont {Bertsch}\ \emph {et~al.}(1984)\citenamefont
  {Bertsch}, \citenamefont {Kruse},\ and\ \citenamefont
  {Gupta}}]{Bertsch:1984gb}%
  \BibitemOpen
  \bibfield  {author} {\bibinfo {author} {\bibfnamefont {G.}~\bibnamefont
  {Bertsch}}, \bibinfo {author} {\bibfnamefont {H.}~\bibnamefont {Kruse}}, \
  and\ \bibinfo {author} {\bibfnamefont {S.}~\bibnamefont {Gupta}},\ }\href
  {\doibase 10.1103/PhysRevC.33.1107} {\bibfield  {journal} {\bibinfo
  {journal} {Phys. Rev. C}\ }\textbf {\bibinfo {volume} {29}},\ \bibinfo
  {pages} {673} (\bibinfo {year} {1984})},\ \bibinfo {note} {[Erratum:
  Phys.Rev.C 33, 1107--1108 (1986)]}\BibitemShut {NoStop}%
\bibitem [{\citenamefont {Stoecker}\ and\ \citenamefont
  {Greiner}(1986)}]{Stoecker:1986ci}%
  \BibitemOpen
  \bibfield  {author} {\bibinfo {author} {\bibfnamefont {H.}~\bibnamefont
  {Stoecker}}\ and\ \bibinfo {author} {\bibfnamefont {W.}~\bibnamefont
  {Greiner}},\ }\href {\doibase 10.1016/0370-1573(86)90131-6} {\bibfield
  {journal} {\bibinfo  {journal} {Phys. Rept.}\ }\textbf {\bibinfo {volume}
  {137}},\ \bibinfo {pages} {277} (\bibinfo {year} {1986})}\BibitemShut
  {NoStop}%
\bibitem [{\citenamefont {Bauer}\ \emph {et~al.}(1986)\citenamefont {Bauer},
  \citenamefont {Bertsch}, \citenamefont {Cassing},\ and\ \citenamefont
  {Mosel}}]{Bauer:1986zz}%
  \BibitemOpen
  \bibfield  {author} {\bibinfo {author} {\bibfnamefont {W.}~\bibnamefont
  {Bauer}}, \bibinfo {author} {\bibfnamefont {G.}~\bibnamefont {Bertsch}},
  \bibinfo {author} {\bibfnamefont {W.}~\bibnamefont {Cassing}}, \ and\
  \bibinfo {author} {\bibfnamefont {U.}~\bibnamefont {Mosel}},\ }\href
  {\doibase 10.1103/PhysRevC.34.2127} {\bibfield  {journal} {\bibinfo
  {journal} {Phys. Rev. C}\ }\textbf {\bibinfo {volume} {34}},\ \bibinfo
  {pages} {2127} (\bibinfo {year} {1986})}\BibitemShut {NoStop}%
\bibitem [{\citenamefont {Bertsch}\ and\ \citenamefont
  {Das~Gupta}(1988)}]{Bertsch:1988ik}%
  \BibitemOpen
  \bibfield  {author} {\bibinfo {author} {\bibfnamefont {G.}~\bibnamefont
  {Bertsch}}\ and\ \bibinfo {author} {\bibfnamefont {S.}~\bibnamefont
  {Das~Gupta}},\ }\href {\doibase 10.1016/0370-1573(88)90170-6} {\bibfield
  {journal} {\bibinfo  {journal} {Phys. Rept.}\ }\textbf {\bibinfo {volume}
  {160}},\ \bibinfo {pages} {189} (\bibinfo {year} {1988})}\BibitemShut
  {NoStop}%
\bibitem [{\citenamefont {Danielewicz}\ and\ \citenamefont
  {Bertsch}(1991)}]{Danielewicz:1991dh}%
  \BibitemOpen
  \bibfield  {author} {\bibinfo {author} {\bibfnamefont {P.}~\bibnamefont
  {Danielewicz}}\ and\ \bibinfo {author} {\bibfnamefont {G.}~\bibnamefont
  {Bertsch}},\ }\href {\doibase 10.1016/0375-9474(91)90541-D} {\bibfield
  {journal} {\bibinfo  {journal} {Nucl. Phys. A}\ }\textbf {\bibinfo {volume}
  {533}},\ \bibinfo {pages} {712} (\bibinfo {year} {1991})}\BibitemShut
  {NoStop}%
\bibitem [{\citenamefont {Kadanoff}\ and\ \citenamefont
  {Baym}(1962)}]{KadanoffBaym:1962}%
  \BibitemOpen
  \bibfield  {author} {\bibinfo {author} {\bibfnamefont {L.}~\bibnamefont
  {Kadanoff}}\ and\ \bibinfo {author} {\bibfnamefont {G.}~\bibnamefont
  {Baym}},\ }\href@noop {} {\emph {\bibinfo {title} {{Quantum Statistical
  Mechanics}}}}\ (\bibinfo  {publisher} {Benjamin, New York},\ \bibinfo {year}
  {1962})\BibitemShut {NoStop}%
\bibitem [{\citenamefont {Botermans}\ and\ \citenamefont
  {Malfliet}(1990)}]{Botermans:1990qi}%
  \BibitemOpen
  \bibfield  {author} {\bibinfo {author} {\bibfnamefont {W.}~\bibnamefont
  {Botermans}}\ and\ \bibinfo {author} {\bibfnamefont {R.}~\bibnamefont
  {Malfliet}},\ }\href {\doibase 10.1016/0370-1573(90)90174-Z} {\bibfield
  {journal} {\bibinfo  {journal} {Phys. Rept.}\ }\textbf {\bibinfo {volume}
  {198}},\ \bibinfo {pages} {115} (\bibinfo {year} {1990})}\BibitemShut
  {NoStop}%
\bibitem [{\citenamefont {Cassing}\ \emph {et~al.}(1990)\citenamefont
  {Cassing}, \citenamefont {Metag}, \citenamefont {Mosel},\ and\ \citenamefont
  {Niita}}]{Cassing:1990dr}%
  \BibitemOpen
  \bibfield  {author} {\bibinfo {author} {\bibfnamefont {W.}~\bibnamefont
  {Cassing}}, \bibinfo {author} {\bibfnamefont {V.}~\bibnamefont {Metag}},
  \bibinfo {author} {\bibfnamefont {U.}~\bibnamefont {Mosel}}, \ and\ \bibinfo
  {author} {\bibfnamefont {K.}~\bibnamefont {Niita}},\ }\href {\doibase
  10.1016/0370-1573(90)90164-W} {\bibfield  {journal} {\bibinfo  {journal}
  {Phys. Rept.}\ }\textbf {\bibinfo {volume} {188}},\ \bibinfo {pages} {363}
  (\bibinfo {year} {1990})}\BibitemShut {NoStop}%
\bibitem [{\citenamefont {Teis}\ \emph {et~al.}(1997)\citenamefont {Teis},
  \citenamefont {Cassing}, \citenamefont {Effenberger}, \citenamefont
  {Hombach}, \citenamefont {Mosel},\ and\ \citenamefont {Wolf}}]{Teis:1996kx}%
  \BibitemOpen
  \bibfield  {author} {\bibinfo {author} {\bibfnamefont {S.}~\bibnamefont
  {Teis}}, \bibinfo {author} {\bibfnamefont {W.}~\bibnamefont {Cassing}},
  \bibinfo {author} {\bibfnamefont {M.}~\bibnamefont {Effenberger}}, \bibinfo
  {author} {\bibfnamefont {A.}~\bibnamefont {Hombach}}, \bibinfo {author}
  {\bibfnamefont {U.}~\bibnamefont {Mosel}}, \ and\ \bibinfo {author}
  {\bibfnamefont {G.}~\bibnamefont {Wolf}},\ }\href {\doibase
  10.1007/s002180050198} {\bibfield  {journal} {\bibinfo  {journal} {Z. Phys.
  A}\ }\textbf {\bibinfo {volume} {356}},\ \bibinfo {pages} {421} (\bibinfo
  {year} {1997})},\ \Eprint {http://arxiv.org/abs/nucl-th/9609009}
  {arXiv:nucl-th/9609009} \BibitemShut {NoStop}%
\bibitem [{\citenamefont {Buss}\ \emph {et~al.}(2012)\citenamefont {Buss},
  \citenamefont {Gaitanos}, \citenamefont {Gallmeister}, \citenamefont {van
  Hees}, \citenamefont {Kaskulov}, \citenamefont {Lalakulich}, \citenamefont
  {Larionov}, \citenamefont {Leitner}, \citenamefont {Weil},\ and\
  \citenamefont {Mosel}}]{Buss:2011mx}%
  \BibitemOpen
  \bibfield  {author} {\bibinfo {author} {\bibfnamefont {O.}~\bibnamefont
  {Buss}}, \bibinfo {author} {\bibfnamefont {T.}~\bibnamefont {Gaitanos}},
  \bibinfo {author} {\bibfnamefont {K.}~\bibnamefont {Gallmeister}}, \bibinfo
  {author} {\bibfnamefont {H.}~\bibnamefont {van Hees}}, \bibinfo {author}
  {\bibfnamefont {M.}~\bibnamefont {Kaskulov}}, \bibinfo {author}
  {\bibfnamefont {O.}~\bibnamefont {Lalakulich}}, \bibinfo {author}
  {\bibfnamefont {A.~B.}\ \bibnamefont {Larionov}}, \bibinfo {author}
  {\bibfnamefont {T.}~\bibnamefont {Leitner}}, \bibinfo {author} {\bibfnamefont
  {J.}~\bibnamefont {Weil}}, \ and\ \bibinfo {author} {\bibfnamefont
  {U.}~\bibnamefont {Mosel}},\ }\href {\doibase 10.1016/j.physrep.2011.12.001}
  {\bibfield  {journal} {\bibinfo  {journal} {Phys. Rept.}\ }\textbf {\bibinfo
  {volume} {512}},\ \bibinfo {pages} {1} (\bibinfo {year} {2012})},\ \Eprint
  {http://arxiv.org/abs/1106.1344} {arXiv:1106.1344 [hep-ph]} \BibitemShut
  {NoStop}%
\bibitem [{\citenamefont {Mosel}(2019)}]{Mosel:2019vhx}%
  \BibitemOpen
  \bibfield  {author} {\bibinfo {author} {\bibfnamefont {U.}~\bibnamefont
  {Mosel}},\ }\href {\doibase 10.1088/1361-6471/ab3830} {\bibfield  {journal}
  {\bibinfo  {journal} {J. Phys. G}\ }\textbf {\bibinfo {volume} {46}},\
  \bibinfo {pages} {113001} (\bibinfo {year} {2019})},\ \Eprint
  {http://arxiv.org/abs/1904.11506} {arXiv:1904.11506 [hep-ex]} \BibitemShut
  {NoStop}%
\bibitem [{\citenamefont {Cugnon}\ \emph {et~al.}(1997)\citenamefont {Cugnon},
  \citenamefont {Volant},\ and\ \citenamefont {Vuillier}}]{Cugnon:1996xf}%
  \BibitemOpen
  \bibfield  {author} {\bibinfo {author} {\bibfnamefont {J.}~\bibnamefont
  {Cugnon}}, \bibinfo {author} {\bibfnamefont {C.}~\bibnamefont {Volant}}, \
  and\ \bibinfo {author} {\bibfnamefont {S.}~\bibnamefont {Vuillier}},\ }\href
  {\doibase 10.1016/S0375-9474(97)00186-3} {\bibfield  {journal} {\bibinfo
  {journal} {Nucl. Phys. A}\ }\textbf {\bibinfo {volume} {620}},\ \bibinfo
  {pages} {475} (\bibinfo {year} {1997})}\BibitemShut {NoStop}%
\bibitem [{\citenamefont {Duarte}(2007)}]{Duarte:2007jd}%
  \BibitemOpen
  \bibfield  {author} {\bibinfo {author} {\bibfnamefont {H.}~\bibnamefont
  {Duarte}},\ }\href {\doibase 10.1103/PhysRevC.75.024611} {\bibfield
  {journal} {\bibinfo  {journal} {Phys. Rev. C}\ }\textbf {\bibinfo {volume}
  {75}},\ \bibinfo {pages} {024611} (\bibinfo {year} {2007})}\BibitemShut
  {NoStop}%
\bibitem [{\citenamefont {Iwamoto}\ \emph {et~al.}(2010)\citenamefont
  {Iwamoto}, \citenamefont {Imamura}, \citenamefont {Koba}, \citenamefont
  {Fukui}, \citenamefont {Wakabayashi}, \citenamefont {Uozumi}, \citenamefont
  {Kin}, \citenamefont {Iwamoto}, \citenamefont {Hohara},\ and\ \citenamefont
  {Nakano}}]{Iwamoto:2010zzb}%
  \BibitemOpen
  \bibfield  {author} {\bibinfo {author} {\bibfnamefont {H.}~\bibnamefont
  {Iwamoto}}, \bibinfo {author} {\bibfnamefont {M.}~\bibnamefont {Imamura}},
  \bibinfo {author} {\bibfnamefont {Y.}~\bibnamefont {Koba}}, \bibinfo {author}
  {\bibfnamefont {Y.}~\bibnamefont {Fukui}}, \bibinfo {author} {\bibfnamefont
  {G.}~\bibnamefont {Wakabayashi}}, \bibinfo {author} {\bibfnamefont
  {Y.}~\bibnamefont {Uozumi}}, \bibinfo {author} {\bibfnamefont
  {T.}~\bibnamefont {Kin}}, \bibinfo {author} {\bibfnamefont {Y.}~\bibnamefont
  {Iwamoto}}, \bibinfo {author} {\bibfnamefont {S.}~\bibnamefont {Hohara}}, \
  and\ \bibinfo {author} {\bibfnamefont {M.}~\bibnamefont {Nakano}},\ }\href
  {\doibase 10.1103/PhysRevC.82.034604} {\bibfield  {journal} {\bibinfo
  {journal} {Phys. Rev. C}\ }\textbf {\bibinfo {volume} {82}},\ \bibinfo
  {pages} {034604} (\bibinfo {year} {2010})}\BibitemShut {NoStop}%
\bibitem [{\citenamefont {Sawada}\ \emph {et~al.}(2012)\citenamefont {Sawada},
  \citenamefont {Uozumi}, \citenamefont {Nogamine}, \citenamefont {Yamada},
  \citenamefont {Iwamoto}, \citenamefont {Sato},\ and\ \citenamefont
  {Niita}}]{Sawada:2012hk}%
  \BibitemOpen
  \bibfield  {author} {\bibinfo {author} {\bibfnamefont {Y.}~\bibnamefont
  {Sawada}}, \bibinfo {author} {\bibfnamefont {Y.}~\bibnamefont {Uozumi}},
  \bibinfo {author} {\bibfnamefont {S.}~\bibnamefont {Nogamine}}, \bibinfo
  {author} {\bibfnamefont {T.}~\bibnamefont {Yamada}}, \bibinfo {author}
  {\bibfnamefont {Y.}~\bibnamefont {Iwamoto}}, \bibinfo {author} {\bibfnamefont
  {T.}~\bibnamefont {Sato}}, \ and\ \bibinfo {author} {\bibfnamefont
  {K.}~\bibnamefont {Niita}},\ }\href {\doibase 10.1016/j.nimb.2012.08.025}
  {\bibfield  {journal} {\bibinfo  {journal} {Nucl. Instrum. Meth. B}\ }\textbf
  {\bibinfo {volume} {291}},\ \bibinfo {pages} {38} (\bibinfo {year}
  {2012})}\BibitemShut {NoStop}%
\bibitem [{\citenamefont {Glauber}\ \emph {et~al.}(1959)\citenamefont {Glauber}
  \emph {et~al.}}]{glauber1959lectures}%
  \BibitemOpen
  \bibfield  {author} {\bibinfo {author} {\bibfnamefont {R.}~\bibnamefont
  {Glauber}} \emph {et~al.},\ }\href@noop {} {\  (\bibinfo {year}
  {1959})}\BibitemShut {NoStop}%
\bibitem [{\citenamefont {Golubeva}\ \emph {et~al.}(1998)\citenamefont
  {Golubeva}, \citenamefont {Kondratyuk}, \citenamefont {Bianconi},
  \citenamefont {Boffi},\ and\ \citenamefont {Radici}}]{Golubeva:1997at}%
  \BibitemOpen
  \bibfield  {author} {\bibinfo {author} {\bibfnamefont {Y.}~\bibnamefont
  {Golubeva}}, \bibinfo {author} {\bibfnamefont {L.}~\bibnamefont
  {Kondratyuk}}, \bibinfo {author} {\bibfnamefont {A.}~\bibnamefont
  {Bianconi}}, \bibinfo {author} {\bibfnamefont {S.}~\bibnamefont {Boffi}}, \
  and\ \bibinfo {author} {\bibfnamefont {M.}~\bibnamefont {Radici}},\ }\href
  {\doibase 10.1103/PhysRevC.57.2618} {\bibfield  {journal} {\bibinfo
  {journal} {Phys. Rev. C}\ }\textbf {\bibinfo {volume} {57}},\ \bibinfo
  {pages} {2618} (\bibinfo {year} {1998})},\ \Eprint
  {http://arxiv.org/abs/nucl-th/9712040} {arXiv:nucl-th/9712040} \BibitemShut
  {NoStop}%
\bibitem [{\citenamefont {Uozumi}\ \emph {et~al.}(2012)\citenamefont {Uozumi},
  \citenamefont {Yamada}, \citenamefont {Nogamine},\ and\ \citenamefont
  {Nakano}}]{Uozumi:2012fm}%
  \BibitemOpen
  \bibfield  {author} {\bibinfo {author} {\bibfnamefont {Y.}~\bibnamefont
  {Uozumi}}, \bibinfo {author} {\bibfnamefont {T.}~\bibnamefont {Yamada}},
  \bibinfo {author} {\bibfnamefont {S.}~\bibnamefont {Nogamine}}, \ and\
  \bibinfo {author} {\bibfnamefont {M.}~\bibnamefont {Nakano}},\ }\href
  {\doibase 10.1103/PhysRevC.86.034610} {\bibfield  {journal} {\bibinfo
  {journal} {Phys. Rev. C}\ }\textbf {\bibinfo {volume} {86}},\ \bibinfo
  {pages} {034610} (\bibinfo {year} {2012})}\BibitemShut {NoStop}%
\bibitem [{\citenamefont {Boudard}\ \emph {et~al.}(2002)\citenamefont
  {Boudard}, \citenamefont {Cugnon}, \citenamefont {Leray},\ and\ \citenamefont
  {Volant}}]{Boudard:2002yn}%
  \BibitemOpen
  \bibfield  {author} {\bibinfo {author} {\bibfnamefont {A.}~\bibnamefont
  {Boudard}}, \bibinfo {author} {\bibfnamefont {J.}~\bibnamefont {Cugnon}},
  \bibinfo {author} {\bibfnamefont {S.}~\bibnamefont {Leray}}, \ and\ \bibinfo
  {author} {\bibfnamefont {C.}~\bibnamefont {Volant}},\ }\href {\doibase
  10.1103/PhysRevC.66.044615} {\bibfield  {journal} {\bibinfo  {journal} {Phys.
  Rev.}\ }\textbf {\bibinfo {volume} {C66}},\ \bibinfo {pages} {044615}
  (\bibinfo {year} {2002})}\BibitemShut {NoStop}%
\bibitem [{\citenamefont {Hayato}(2002)}]{Hayato:2002sd}%
  \BibitemOpen
  \bibfield  {author} {\bibinfo {author} {\bibfnamefont {Y.}~\bibnamefont
  {Hayato}},\ }\href {\doibase 10.1016/S0920-5632(02)01759-0} {\bibfield
  {journal} {\bibinfo  {journal} {Nucl. Phys. B Proc. Suppl.}\ }\textbf
  {\bibinfo {volume} {112}},\ \bibinfo {pages} {171} (\bibinfo {year}
  {2002})}\BibitemShut {NoStop}%
\bibitem [{\citenamefont {Casper}(2002)}]{Casper:2002sd}%
  \BibitemOpen
  \bibfield  {author} {\bibinfo {author} {\bibfnamefont {D.}~\bibnamefont
  {Casper}},\ }\href {\doibase 10.1016/S0920-5632(02)01756-5} {\bibfield
  {journal} {\bibinfo  {journal} {Nucl. Phys. B Proc. Suppl.}\ }\textbf
  {\bibinfo {volume} {112}},\ \bibinfo {pages} {161} (\bibinfo {year}
  {2002})},\ \Eprint {http://arxiv.org/abs/hep-ph/0208030}
  {arXiv:hep-ph/0208030} \BibitemShut {NoStop}%
\bibitem [{\citenamefont {Battistoni}\ \emph {et~al.}(2013)\citenamefont
  {Battistoni}, \citenamefont {Cerutti}, \citenamefont {Ferrari}, \citenamefont
  {Ranft}, \citenamefont {Roesler},\ and\ \citenamefont
  {Sala}}]{Battistoni:2013tra}%
  \BibitemOpen
  \bibfield  {author} {\bibinfo {author} {\bibfnamefont {G.}~\bibnamefont
  {Battistoni}}, \bibinfo {author} {\bibfnamefont {F.}~\bibnamefont {Cerutti}},
  \bibinfo {author} {\bibfnamefont {A.}~\bibnamefont {Ferrari}}, \bibinfo
  {author} {\bibfnamefont {J.}~\bibnamefont {Ranft}}, \bibinfo {author}
  {\bibfnamefont {S.}~\bibnamefont {Roesler}}, \ and\ \bibinfo {author}
  {\bibfnamefont {P.}~\bibnamefont {Sala}},\ }\href {\doibase
  10.1088/1742-6596/408/1/012051} {\bibfield  {journal} {\bibinfo  {journal}
  {J. Phys. Conf. Ser.}\ }\textbf {\bibinfo {volume} {408}},\ \bibinfo {pages}
  {012051} (\bibinfo {year} {2013})}\BibitemShut {NoStop}%
\bibitem [{\citenamefont {Battistoni}\ \emph {et~al.}(2015)\citenamefont
  {Battistoni} \emph {et~al.}}]{Battistoni:2015epi}%
  \BibitemOpen
  \bibfield  {author} {\bibinfo {author} {\bibfnamefont {G.}~\bibnamefont
  {Battistoni}} \emph {et~al.},\ }\href {\doibase
  10.1016/j.anucene.2014.11.007} {\bibfield  {journal} {\bibinfo  {journal}
  {Annals Nucl. Energy}\ }\textbf {\bibinfo {volume} {82}},\ \bibinfo {pages}
  {10} (\bibinfo {year} {2015})}\BibitemShut {NoStop}%
\bibitem [{\citenamefont {Golan}\ \emph {et~al.}(2012)\citenamefont {Golan},
  \citenamefont {Juszczak},\ and\ \citenamefont {Sobczyk}}]{Golan:2012wx}%
  \BibitemOpen
  \bibfield  {author} {\bibinfo {author} {\bibfnamefont {T.}~\bibnamefont
  {Golan}}, \bibinfo {author} {\bibfnamefont {C.}~\bibnamefont {Juszczak}}, \
  and\ \bibinfo {author} {\bibfnamefont {J.~T.}\ \bibnamefont {Sobczyk}},\
  }\href {\doibase 10.1103/PhysRevC.86.015505} {\bibfield  {journal} {\bibinfo
  {journal} {Phys. Rev.}\ }\textbf {\bibinfo {volume} {C86}},\ \bibinfo {pages}
  {015505} (\bibinfo {year} {2012})},\ \Eprint {http://arxiv.org/abs/1202.4197}
  {arXiv:1202.4197 [nucl-th]} \BibitemShut {NoStop}%
\bibitem [{\citenamefont {Andreopoulos}\ \emph {et~al.}(2010)\citenamefont
  {Andreopoulos} \emph {et~al.}}]{Andreopoulos:2009rq}%
  \BibitemOpen
  \bibfield  {author} {\bibinfo {author} {\bibfnamefont {C.}~\bibnamefont
  {Andreopoulos}} \emph {et~al.},\ }\href {\doibase 10.1016/j.nima.2009.12.009}
  {\bibfield  {journal} {\bibinfo  {journal} {Nucl. Instrum. Meth. A}\ }\textbf
  {\bibinfo {volume} {614}},\ \bibinfo {pages} {87} (\bibinfo {year} {2010})},\
  \Eprint {http://arxiv.org/abs/0905.2517} {arXiv:0905.2517 [hep-ph]}
  \BibitemShut {NoStop}%
\bibitem [{\citenamefont {Niewczas}\ and\ \citenamefont
  {Sobczyk}(2019)}]{Niewczas:2019fro}%
  \BibitemOpen
  \bibfield  {author} {\bibinfo {author} {\bibfnamefont {K.}~\bibnamefont
  {Niewczas}}\ and\ \bibinfo {author} {\bibfnamefont {J.~T.}\ \bibnamefont
  {Sobczyk}},\ }\href {\doibase 10.1103/PhysRevC.100.015505} {\bibfield
  {journal} {\bibinfo  {journal} {Phys. Rev. C}\ }\textbf {\bibinfo {volume}
  {100}},\ \bibinfo {pages} {015505} (\bibinfo {year} {2019})},\ \Eprint
  {http://arxiv.org/abs/1902.05618} {arXiv:1902.05618 [hep-ex]} \BibitemShut
  {NoStop}%
\bibitem [{\citenamefont {Carlson}\ \emph {et~al.}(2015)\citenamefont
  {Carlson}, \citenamefont {Gandolfi}, \citenamefont {Pederiva}, \citenamefont
  {Pieper}, \citenamefont {Schiavilla}, \citenamefont {Schmidt},\ and\
  \citenamefont {Wiringa}}]{Carlson:2014vla}%
  \BibitemOpen
  \bibfield  {author} {\bibinfo {author} {\bibfnamefont {J.}~\bibnamefont
  {Carlson}}, \bibinfo {author} {\bibfnamefont {S.}~\bibnamefont {Gandolfi}},
  \bibinfo {author} {\bibfnamefont {F.}~\bibnamefont {Pederiva}}, \bibinfo
  {author} {\bibfnamefont {S.~C.}\ \bibnamefont {Pieper}}, \bibinfo {author}
  {\bibfnamefont {R.}~\bibnamefont {Schiavilla}}, \bibinfo {author}
  {\bibfnamefont {K.}~\bibnamefont {Schmidt}}, \ and\ \bibinfo {author}
  {\bibfnamefont {R.}~\bibnamefont {Wiringa}},\ }\href {\doibase
  10.1103/RevModPhys.87.1067} {\bibfield  {journal} {\bibinfo  {journal} {Rev.
  Mod. Phys.}\ }\textbf {\bibinfo {volume} {87}},\ \bibinfo {pages} {1067}
  (\bibinfo {year} {2015})},\ \Eprint {http://arxiv.org/abs/1412.3081}
  {arXiv:1412.3081 [nucl-th]} \BibitemShut {NoStop}%
\bibitem [{\citenamefont {Wiringa}\ \emph {et~al.}(1995)\citenamefont
  {Wiringa}, \citenamefont {Stoks},\ and\ \citenamefont
  {Schiavilla}}]{Wiringa:1994wb}%
  \BibitemOpen
  \bibfield  {author} {\bibinfo {author} {\bibfnamefont {R.~B.}\ \bibnamefont
  {Wiringa}}, \bibinfo {author} {\bibfnamefont {V.~G.~J.}\ \bibnamefont
  {Stoks}}, \ and\ \bibinfo {author} {\bibfnamefont {R.}~\bibnamefont
  {Schiavilla}},\ }\href {\doibase 10.1103/PhysRevC.51.38} {\bibfield
  {journal} {\bibinfo  {journal} {Phys. Rev.}\ }\textbf {\bibinfo {volume}
  {C51}},\ \bibinfo {pages} {38} (\bibinfo {year} {1995})},\ \Eprint
  {http://arxiv.org/abs/nucl-th/9408016} {arXiv:nucl-th/9408016 [nucl-th]}
  \BibitemShut {NoStop}%
\bibitem [{\citenamefont {Pieper}(2008)}]{Pieper:2008}%
  \BibitemOpen
  \bibfield  {author} {\bibinfo {author} {\bibfnamefont {S.~C.}\ \bibnamefont
  {Pieper}},\ }\href {\doibase 10.1063/1.2932280} {\bibfield  {journal}
  {\bibinfo  {journal} {AIP Conference Proceedings}\ }\textbf {\bibinfo
  {volume} {1011}},\ \bibinfo {pages} {143} (\bibinfo {year} {2008})},\ \Eprint
  {http://arxiv.org/abs/http://aip.scitation.org/doi/pdf/10.1063/1.2932280}
  {http://aip.scitation.org/doi/pdf/10.1063/1.2932280} \BibitemShut {NoStop}%
\bibitem [{\citenamefont {Landau}\ and\ \citenamefont
  {Pomeranchuk}(1953)}]{Landau:1953gr}%
  \BibitemOpen
  \bibfield  {author} {\bibinfo {author} {\bibfnamefont {L.~D.}\ \bibnamefont
  {Landau}}\ and\ \bibinfo {author} {\bibfnamefont {I.}~\bibnamefont
  {Pomeranchuk}},\ }\href@noop {} {\bibfield  {journal} {\bibinfo  {journal}
  {Dokl. Akad. Nauk Ser. Fiz.}\ }\textbf {\bibinfo {volume} {92}},\ \bibinfo
  {pages} {735} (\bibinfo {year} {1953})}\BibitemShut {NoStop}%
\bibitem [{\citenamefont {Stodolsky}(1975)}]{Stodolsky:1975qq}%
  \BibitemOpen
  \bibfield  {author} {\bibinfo {author} {\bibfnamefont {L.}~\bibnamefont
  {Stodolsky}},\ }in\ \href@noop {} {\emph {\bibinfo {booktitle} {{Proceedings:
  International Colloquium on Multiparticle Reactions, 6th, Oxford, Eng., Jul
  14-19, 1975}}}}\ (\bibinfo {year} {1975})\ p.\ \bibinfo {pages}
  {577}\BibitemShut {NoStop}%
\bibitem [{\citenamefont {Bogner}\ \emph {et~al.}(2010)\citenamefont {Bogner},
  \citenamefont {Furnstahl},\ and\ \citenamefont {Schwenk}}]{Bogner:2009bt}%
  \BibitemOpen
  \bibfield  {author} {\bibinfo {author} {\bibfnamefont {S.}~\bibnamefont
  {Bogner}}, \bibinfo {author} {\bibfnamefont {R.}~\bibnamefont {Furnstahl}}, \
  and\ \bibinfo {author} {\bibfnamefont {A.}~\bibnamefont {Schwenk}},\ }\href
  {\doibase 10.1016/j.ppnp.2010.03.001} {\bibfield  {journal} {\bibinfo
  {journal} {Prog. Part. Nucl. Phys.}\ }\textbf {\bibinfo {volume} {65}},\
  \bibinfo {pages} {94} (\bibinfo {year} {2010})},\ \Eprint
  {http://arxiv.org/abs/0912.3688} {arXiv:0912.3688 [nucl-th]} \BibitemShut
  {NoStop}%
\bibitem [{\citenamefont {Barrett}\ \emph {et~al.}(2013)\citenamefont
  {Barrett}, \citenamefont {Navratil},\ and\ \citenamefont
  {Vary}}]{Barrett:2013nh}%
  \BibitemOpen
  \bibfield  {author} {\bibinfo {author} {\bibfnamefont {B.~R.}\ \bibnamefont
  {Barrett}}, \bibinfo {author} {\bibfnamefont {P.}~\bibnamefont {Navratil}}, \
  and\ \bibinfo {author} {\bibfnamefont {J.~P.}\ \bibnamefont {Vary}},\ }\href
  {\doibase 10.1016/j.ppnp.2012.10.003} {\bibfield  {journal} {\bibinfo
  {journal} {Prog. Part. Nucl. Phys.}\ }\textbf {\bibinfo {volume} {69}},\
  \bibinfo {pages} {131} (\bibinfo {year} {2013})}\BibitemShut {NoStop}%
\bibitem [{\citenamefont {Hagen}\ \emph {et~al.}(2014)\citenamefont {Hagen},
  \citenamefont {Papenbrock}, \citenamefont {Hjorth-Jensen},\ and\
  \citenamefont {Dean}}]{Hagen:2013nca}%
  \BibitemOpen
  \bibfield  {author} {\bibinfo {author} {\bibfnamefont {G.}~\bibnamefont
  {Hagen}}, \bibinfo {author} {\bibfnamefont {T.}~\bibnamefont {Papenbrock}},
  \bibinfo {author} {\bibfnamefont {M.}~\bibnamefont {Hjorth-Jensen}}, \ and\
  \bibinfo {author} {\bibfnamefont {D.}~\bibnamefont {Dean}},\ }\href {\doibase
  10.1088/0034-4885/77/9/096302} {\bibfield  {journal} {\bibinfo  {journal}
  {Rept. Prog. Phys.}\ }\textbf {\bibinfo {volume} {77}},\ \bibinfo {pages}
  {096302} (\bibinfo {year} {2014})},\ \Eprint {http://arxiv.org/abs/1312.7872}
  {arXiv:1312.7872 [nucl-th]} \BibitemShut {NoStop}%
\bibitem [{\citenamefont {Barbieri}\ and\ \citenamefont
  {Carbone}(2017)}]{Barbieri:2016uib}%
  \BibitemOpen
  \bibfield  {author} {\bibinfo {author} {\bibfnamefont {C.}~\bibnamefont
  {Barbieri}}\ and\ \bibinfo {author} {\bibfnamefont {A.}~\bibnamefont
  {Carbone}},\ }\enquote {\bibinfo {title} {{Self-consistent Green's function
  approaches}},}\ \ (\bibinfo {year} {2017})\ pp.\ \bibinfo {pages}
  {571--644},\ \Eprint {http://arxiv.org/abs/1611.03923} {arXiv:1611.03923
  [nucl-th]} \BibitemShut {NoStop}%
\bibitem [{\citenamefont {Wiringa}\ \emph {et~al.}(2000)\citenamefont
  {Wiringa}, \citenamefont {Pieper}, \citenamefont {Carlson},\ and\
  \citenamefont {Pandharipande}}]{Wiringa:2000gb}%
  \BibitemOpen
  \bibfield  {author} {\bibinfo {author} {\bibfnamefont {R.~B.}\ \bibnamefont
  {Wiringa}}, \bibinfo {author} {\bibfnamefont {S.~C.}\ \bibnamefont {Pieper}},
  \bibinfo {author} {\bibfnamefont {J.}~\bibnamefont {Carlson}}, \ and\
  \bibinfo {author} {\bibfnamefont {V.}~\bibnamefont {Pandharipande}},\ }\href
  {\doibase 10.1103/PhysRevC.62.014001} {\bibfield  {journal} {\bibinfo
  {journal} {Phys. Rev. C}\ }\textbf {\bibinfo {volume} {62}},\ \bibinfo
  {pages} {014001} (\bibinfo {year} {2000})},\ \Eprint
  {http://arxiv.org/abs/nucl-th/0002022} {arXiv:nucl-th/0002022} \BibitemShut
  {NoStop}%
\bibitem [{\citenamefont {Lovato}\ \emph {et~al.}(2013)\citenamefont {Lovato},
  \citenamefont {Gandolfi}, \citenamefont {Butler}, \citenamefont {Carlson},
  \citenamefont {Lusk}, \citenamefont {Pieper},\ and\ \citenamefont
  {Schiavilla}}]{Lovato:2013cua}%
  \BibitemOpen
  \bibfield  {author} {\bibinfo {author} {\bibfnamefont {A.}~\bibnamefont
  {Lovato}}, \bibinfo {author} {\bibfnamefont {S.}~\bibnamefont {Gandolfi}},
  \bibinfo {author} {\bibfnamefont {R.}~\bibnamefont {Butler}}, \bibinfo
  {author} {\bibfnamefont {J.}~\bibnamefont {Carlson}}, \bibinfo {author}
  {\bibfnamefont {E.}~\bibnamefont {Lusk}}, \bibinfo {author} {\bibfnamefont
  {S.~C.}\ \bibnamefont {Pieper}}, \ and\ \bibinfo {author} {\bibfnamefont
  {R.}~\bibnamefont {Schiavilla}},\ }\href {\doibase
  10.1103/PhysRevLett.111.092501} {\bibfield  {journal} {\bibinfo  {journal}
  {Phys. Rev. Lett.}\ }\textbf {\bibinfo {volume} {111}},\ \bibinfo {pages}
  {092501} (\bibinfo {year} {2013})},\ \Eprint {http://arxiv.org/abs/1305.6959}
  {arXiv:1305.6959 [nucl-th]} \BibitemShut {NoStop}%
\bibitem [{\citenamefont {Cruz-Torres}\ \emph {et~al.}(2019)\citenamefont
  {Cruz-Torres}, \citenamefont {Lonardoni}, \citenamefont {Weiss},
  \citenamefont {Barnea}, \citenamefont {Higinbotham}, \citenamefont
  {Piasetzky}, \citenamefont {Schmidt}, \citenamefont {Weinstein},
  \citenamefont {Wiringa},\ and\ \citenamefont {Hen}}]{Cruz-Torres:2019fum}%
  \BibitemOpen
  \bibfield  {author} {\bibinfo {author} {\bibfnamefont {R.}~\bibnamefont
  {Cruz-Torres}}, \bibinfo {author} {\bibfnamefont {D.}~\bibnamefont
  {Lonardoni}}, \bibinfo {author} {\bibfnamefont {R.}~\bibnamefont {Weiss}},
  \bibinfo {author} {\bibfnamefont {N.}~\bibnamefont {Barnea}}, \bibinfo
  {author} {\bibfnamefont {D.}~\bibnamefont {Higinbotham}}, \bibinfo {author}
  {\bibfnamefont {E.}~\bibnamefont {Piasetzky}}, \bibinfo {author}
  {\bibfnamefont {A.}~\bibnamefont {Schmidt}}, \bibinfo {author} {\bibfnamefont
  {L.}~\bibnamefont {Weinstein}}, \bibinfo {author} {\bibfnamefont
  {R.}~\bibnamefont {Wiringa}}, \ and\ \bibinfo {author} {\bibfnamefont
  {O.}~\bibnamefont {Hen}},\ }\href@noop {} {\  (\bibinfo {year} {2019})},\
  \Eprint {http://arxiv.org/abs/1907.03658} {arXiv:1907.03658 [nucl-th]}
  \BibitemShut {NoStop}%
\bibitem [{\citenamefont {Sjostrand}\ and\ \citenamefont {van
  Zijl}(1987)}]{Sjostrand:1987su}%
  \BibitemOpen
  \bibfield  {author} {\bibinfo {author} {\bibfnamefont {T.}~\bibnamefont
  {Sjostrand}}\ and\ \bibinfo {author} {\bibfnamefont {M.}~\bibnamefont {van
  Zijl}},\ }\href {\doibase 10.1103/PhysRevD.36.2019} {\bibfield  {journal}
  {\bibinfo  {journal} {Phys. Rev.}\ }\textbf {\bibinfo {volume} {D36}},\
  \bibinfo {pages} {2019} (\bibinfo {year} {1987})}\BibitemShut {NoStop}%
\bibitem [{\citenamefont {{SAID~database}}()}]{SAID}%
  \BibitemOpen
  \bibfield  {author} {\bibinfo {author} {\bibnamefont {{SAID~database}}},\
  }\href@noop {} {}\bibinfo {howpublished}
  {\url{http://gwdac.phys.gwu.edu/}}\BibitemShut {NoStop}%
\bibitem [{\citenamefont {Agostinelli}\ \emph {et~al.}(2003)\citenamefont
  {Agostinelli} \emph {et~al.}}]{Agostinelli:2002hh}%
  \BibitemOpen
  \bibfield  {author} {\bibinfo {author} {\bibfnamefont {S.}~\bibnamefont
  {Agostinelli}} \emph {et~al.} (\bibinfo {collaboration} {GEANT4}),\ }\href
  {\doibase 10.1016/S0168-9002(03)01368-8} {\bibfield  {journal} {\bibinfo
  {journal} {Nucl. Instrum. Meth. A}\ }\textbf {\bibinfo {volume} {506}},\
  \bibinfo {pages} {250} (\bibinfo {year} {2003})}\BibitemShut {NoStop}%
\bibitem [{\citenamefont {Norbury}(2008)}]{Norbury:2008}%
  \BibitemOpen
  \bibfield  {author} {\bibinfo {author} {\bibfnamefont {J.}~\bibnamefont
  {Norbury}},\ }\href@noop {} {\bibfield  {journal} {\bibinfo  {journal} {NASA
  Technical Paper}\ }\textbf {\bibinfo {volume} {215116}} (\bibinfo {year}
  {2008})}\BibitemShut {NoStop}%
\bibitem [{\citenamefont {Dicello}\ and\ \citenamefont
  {Igo}(1970)}]{Dicello:1970mx}%
  \BibitemOpen
  \bibfield  {author} {\bibinfo {author} {\bibfnamefont {J.~F.}\ \bibnamefont
  {Dicello}}\ and\ \bibinfo {author} {\bibfnamefont {G.}~\bibnamefont {Igo}},\
  }\href {\doibase 10.1103/PhysRevC.2.488} {\bibfield  {journal} {\bibinfo
  {journal} {Phys. Rev.}\ }\textbf {\bibinfo {volume} {C2}},\ \bibinfo {pages}
  {488} (\bibinfo {year} {1970})}\BibitemShut {NoStop}%
\bibitem [{\citenamefont {Bobchenko}\ \emph {et~al.}(1979)\citenamefont
  {Bobchenko} \emph {et~al.}}]{Bobchenko:1979hp}%
  \BibitemOpen
  \bibfield  {author} {\bibinfo {author} {\bibfnamefont {B.~M.}\ \bibnamefont
  {Bobchenko}} \emph {et~al.},\ }\href@noop {} {\bibfield  {journal} {\bibinfo
  {journal} {Sov. J. Nucl. Phys.}\ }\textbf {\bibinfo {volume} {30}},\ \bibinfo
  {pages} {805} (\bibinfo {year} {1979})},\ \bibinfo {note} {[Yad.
  Fiz.30,1553(1979)]}\BibitemShut {NoStop}%
\bibitem [{\citenamefont {Bauhoff}(1986)}]{Bauhoff:1986gcb}%
  \BibitemOpen
  \bibfield  {author} {\bibinfo {author} {\bibfnamefont {W.}~\bibnamefont
  {Bauhoff}},\ }\href {\doibase 10.1016/0092-640X(86)90016-1} {\bibfield
  {journal} {\bibinfo  {journal} {Atom. Data Nucl. Data Tabl.}\ }\textbf
  {\bibinfo {volume} {35}},\ \bibinfo {pages} {429} (\bibinfo {year}
  {1986})}\BibitemShut {NoStop}%
\bibitem [{\citenamefont {Kox}\ \emph {et~al.}(1987)\citenamefont {Kox} \emph
  {et~al.}}]{Kox:1987qw}%
  \BibitemOpen
  \bibfield  {author} {\bibinfo {author} {\bibfnamefont {S.}~\bibnamefont
  {Kox}} \emph {et~al.},\ }\href {\doibase 10.1103/PhysRevC.35.1678} {\bibfield
   {journal} {\bibinfo  {journal} {Phys. Rev.}\ }\textbf {\bibinfo {volume}
  {C35}},\ \bibinfo {pages} {1678} (\bibinfo {year} {1987})}\BibitemShut
  {NoStop}%
\bibitem [{\citenamefont {Sihver}\ \emph {et~al.}(1993)\citenamefont {Sihver},
  \citenamefont {Tsao}, \citenamefont {Silberberg}, \citenamefont {Kanai},\
  and\ \citenamefont {Barghouty}}]{Sihver:1993pc}%
  \BibitemOpen
  \bibfield  {author} {\bibinfo {author} {\bibfnamefont {L.}~\bibnamefont
  {Sihver}}, \bibinfo {author} {\bibfnamefont {C.~H.}\ \bibnamefont {Tsao}},
  \bibinfo {author} {\bibfnamefont {R.}~\bibnamefont {Silberberg}}, \bibinfo
  {author} {\bibfnamefont {T.}~\bibnamefont {Kanai}}, \ and\ \bibinfo {author}
  {\bibfnamefont {A.~F.}\ \bibnamefont {Barghouty}},\ }\href {\doibase
  10.1103/PhysRevC.47.1225} {\bibfield  {journal} {\bibinfo  {journal} {Phys.
  Rev.}\ }\textbf {\bibinfo {volume} {C47}},\ \bibinfo {pages} {1225} (\bibinfo
  {year} {1993})}\BibitemShut {NoStop}%
\bibitem [{\citenamefont {Carlson}(1996)}]{Carlson:1996ofz}%
  \BibitemOpen
  \bibfield  {author} {\bibinfo {author} {\bibfnamefont {R.~F.}\ \bibnamefont
  {Carlson}},\ }\href {\doibase 10.1006/adnd.1996.0010} {\bibfield  {journal}
  {\bibinfo  {journal} {Atom. Data Nucl. Data Tabl.}\ }\textbf {\bibinfo
  {volume} {63}},\ \bibinfo {pages} {93} (\bibinfo {year} {1996})}\BibitemShut
  {NoStop}%
\bibitem [{\citenamefont {Pandharipande}\ and\ \citenamefont
  {Pieper}(1992)}]{Pandharipande:1992zz}%
  \BibitemOpen
  \bibfield  {author} {\bibinfo {author} {\bibfnamefont {V.}~\bibnamefont
  {Pandharipande}}\ and\ \bibinfo {author} {\bibfnamefont {S.~C.}\ \bibnamefont
  {Pieper}},\ }\href {\doibase 10.1103/PhysRevC.45.791} {\bibfield  {journal}
  {\bibinfo  {journal} {Phys. Rev. C}\ }\textbf {\bibinfo {volume} {45}},\
  \bibinfo {pages} {791} (\bibinfo {year} {1992})}\BibitemShut {NoStop}%
\bibitem [{\citenamefont {González-Jiménez}\ \emph
  {et~al.}(2019)\citenamefont {González-Jiménez}, \citenamefont
  {Nikolakopoulos}, \citenamefont {Jachowicz},\ and\ \citenamefont
  {Udías}}]{Gonzalez-Jimenez:2019qhq}%
  \BibitemOpen
  \bibfield  {author} {\bibinfo {author} {\bibfnamefont {R.}~\bibnamefont
  {González-Jiménez}}, \bibinfo {author} {\bibfnamefont {A.}~\bibnamefont
  {Nikolakopoulos}}, \bibinfo {author} {\bibfnamefont {N.}~\bibnamefont
  {Jachowicz}}, \ and\ \bibinfo {author} {\bibfnamefont {J.}~\bibnamefont
  {Udías}},\ }\href {\doibase 10.1103/PhysRevC.100.045501} {\bibfield
  {journal} {\bibinfo  {journal} {Phys. Rev. C}\ }\textbf {\bibinfo {volume}
  {100}},\ \bibinfo {pages} {045501} (\bibinfo {year} {2019})},\ \Eprint
  {http://arxiv.org/abs/1904.10696} {arXiv:1904.10696 [nucl-th]} \BibitemShut
  {NoStop}%
\bibitem [{\citenamefont {O'Neill}\ \emph {et~al.}(1995)\citenamefont {O'Neill}
  \emph {et~al.}}]{ONeill:1994znv}%
  \BibitemOpen
  \bibfield  {author} {\bibinfo {author} {\bibfnamefont {T.~G.}\ \bibnamefont
  {O'Neill}} \emph {et~al.},\ }\href {\doibase 10.1016/0370-2693(95)00362-O}
  {\bibfield  {journal} {\bibinfo  {journal} {Phys. Lett.}\ }\textbf {\bibinfo
  {volume} {B351}},\ \bibinfo {pages} {87} (\bibinfo {year} {1995})},\ \Eprint
  {http://arxiv.org/abs/hep-ph/9408260} {arXiv:hep-ph/9408260 [hep-ph]}
  \BibitemShut {NoStop}%
\bibitem [{\citenamefont {Garrow}\ \emph {et~al.}(2002)\citenamefont {Garrow}
  \emph {et~al.}}]{Garrow:2001di}%
  \BibitemOpen
  \bibfield  {author} {\bibinfo {author} {\bibfnamefont {K.}~\bibnamefont
  {Garrow}} \emph {et~al.},\ }\href {\doibase 10.1103/PhysRevC.66.044613}
  {\bibfield  {journal} {\bibinfo  {journal} {Phys. Rev. C}\ }\textbf {\bibinfo
  {volume} {66}},\ \bibinfo {pages} {044613} (\bibinfo {year} {2002})},\
  \Eprint {http://arxiv.org/abs/hep-ex/0109027} {arXiv:hep-ex/0109027}
  \BibitemShut {NoStop}%
\bibitem [{\citenamefont {Garino}\ \emph {et~al.}(1992)\citenamefont {Garino}
  \emph {et~al.}}]{Garino:1992ca}%
  \BibitemOpen
  \bibfield  {author} {\bibinfo {author} {\bibfnamefont {G.}~\bibnamefont
  {Garino}} \emph {et~al.},\ }\href {\doibase 10.1103/PhysRevC.45.780}
  {\bibfield  {journal} {\bibinfo  {journal} {Phys. Rev. C}\ }\textbf {\bibinfo
  {volume} {45}},\ \bibinfo {pages} {780} (\bibinfo {year} {1992})}\BibitemShut
  {NoStop}%
\bibitem [{\citenamefont {Radici}\ \emph {et~al.}(1994)\citenamefont {Radici},
  \citenamefont {Boffi}, \citenamefont {Pieper},\ and\ \citenamefont
  {Pandharipande}}]{Radici:1994wy}%
  \BibitemOpen
  \bibfield  {author} {\bibinfo {author} {\bibfnamefont {M.}~\bibnamefont
  {Radici}}, \bibinfo {author} {\bibfnamefont {S.}~\bibnamefont {Boffi}},
  \bibinfo {author} {\bibfnamefont {S.~C.}\ \bibnamefont {Pieper}}, \ and\
  \bibinfo {author} {\bibfnamefont {V.}~\bibnamefont {Pandharipande}},\ }\href
  {\doibase 10.1103/PhysRevC.50.3010} {\bibfield  {journal} {\bibinfo
  {journal} {Phys. Rev. C}\ }\textbf {\bibinfo {volume} {50}},\ \bibinfo
  {pages} {3010} (\bibinfo {year} {1994})},\ \Eprint
  {http://arxiv.org/abs/nucl-th/9408015} {arXiv:nucl-th/9408015} \BibitemShut
  {NoStop}%
\bibitem [{\citenamefont {Benhar}\ \emph {et~al.}(1993)\citenamefont {Benhar},
  \citenamefont {Pandharipande},\ and\ \citenamefont {Pieper}}]{Benhar:1993zz}%
  \BibitemOpen
  \bibfield  {author} {\bibinfo {author} {\bibfnamefont {O.}~\bibnamefont
  {Benhar}}, \bibinfo {author} {\bibfnamefont {V.}~\bibnamefont
  {Pandharipande}}, \ and\ \bibinfo {author} {\bibfnamefont {S.~C.}\
  \bibnamefont {Pieper}},\ }\href {\doibase 10.1103/RevModPhys.65.817}
  {\bibfield  {journal} {\bibinfo  {journal} {Rev. Mod. Phys.}\ }\textbf
  {\bibinfo {volume} {65}},\ \bibinfo {pages} {817} (\bibinfo {year}
  {1993})}\BibitemShut {NoStop}%
\bibitem [{\citenamefont {Nikolaev}\ \emph {et~al.}(1993)\citenamefont
  {Nikolaev}, \citenamefont {Szczurek}, \citenamefont {Speth}, \citenamefont
  {Wambach}, \citenamefont {Zakharov},\ and\ \citenamefont
  {Zoller}}]{Nikolaev:1993sj}%
  \BibitemOpen
  \bibfield  {author} {\bibinfo {author} {\bibfnamefont {N.~N.}\ \bibnamefont
  {Nikolaev}}, \bibinfo {author} {\bibfnamefont {A.}~\bibnamefont {Szczurek}},
  \bibinfo {author} {\bibfnamefont {J.}~\bibnamefont {Speth}}, \bibinfo
  {author} {\bibfnamefont {J.}~\bibnamefont {Wambach}}, \bibinfo {author}
  {\bibfnamefont {B.}~\bibnamefont {Zakharov}}, \ and\ \bibinfo {author}
  {\bibfnamefont {V.}~\bibnamefont {Zoller}},\ }\href {\doibase
  10.1016/0370-2693(93)90996-U} {\bibfield  {journal} {\bibinfo  {journal}
  {Phys. Lett. B}\ }\textbf {\bibinfo {volume} {317}},\ \bibinfo {pages} {281}
  (\bibinfo {year} {1993})},\ \Eprint {http://arxiv.org/abs/nucl-th/9306007}
  {arXiv:nucl-th/9306007} \BibitemShut {NoStop}%
\bibitem [{\citenamefont {Benhar}(2013)}]{Benhar:2013dq}%
  \BibitemOpen
  \bibfield  {author} {\bibinfo {author} {\bibfnamefont {O.}~\bibnamefont
  {Benhar}},\ }\href {\doibase 10.1103/PhysRevC.87.024606} {\bibfield
  {journal} {\bibinfo  {journal} {Phys. Rev. C}\ }\textbf {\bibinfo {volume}
  {87}},\ \bibinfo {pages} {024606} (\bibinfo {year} {2013})},\ \Eprint
  {http://arxiv.org/abs/1301.3357} {arXiv:1301.3357 [nucl-th]} \BibitemShut
  {NoStop}%
\bibitem [{\citenamefont {Benhar}\ \emph {et~al.}(1995)\citenamefont {Benhar},
  \citenamefont {Fabrocini}, \citenamefont {Fantoni}, \citenamefont {Pieper},
  \citenamefont {Pandharipande},\ and\ \citenamefont {Sick}}]{Benhar:1995xa}%
  \BibitemOpen
  \bibfield  {author} {\bibinfo {author} {\bibfnamefont {O.}~\bibnamefont
  {Benhar}}, \bibinfo {author} {\bibfnamefont {A.}~\bibnamefont {Fabrocini}},
  \bibinfo {author} {\bibfnamefont {S.}~\bibnamefont {Fantoni}}, \bibinfo
  {author} {\bibfnamefont {S.}~\bibnamefont {Pieper}}, \bibinfo {author}
  {\bibfnamefont {V.}~\bibnamefont {Pandharipande}}, \ and\ \bibinfo {author}
  {\bibfnamefont {I.}~\bibnamefont {Sick}},\ }\href {\doibase
  10.1016/0370-2693(95)01027-N} {\bibfield  {journal} {\bibinfo  {journal}
  {Phys. Lett. B}\ }\textbf {\bibinfo {volume} {359}},\ \bibinfo {pages} {8}
  (\bibinfo {year} {1995})}\BibitemShut {NoStop}%
\bibitem [{\citenamefont {Salcedo}\ \emph {et~al.}(1988)\citenamefont
  {Salcedo}, \citenamefont {Oset}, \citenamefont {Vicente-Vacas},\ and\
  \citenamefont {Garcia-Recio}}]{Salcedo:1987md}%
  \BibitemOpen
  \bibfield  {author} {\bibinfo {author} {\bibfnamefont {L.}~\bibnamefont
  {Salcedo}}, \bibinfo {author} {\bibfnamefont {E.}~\bibnamefont {Oset}},
  \bibinfo {author} {\bibfnamefont {M.}~\bibnamefont {Vicente-Vacas}}, \ and\
  \bibinfo {author} {\bibfnamefont {C.}~\bibnamefont {Garcia-Recio}},\ }\href
  {\doibase 10.1016/0375-9474(88)90310-7} {\bibfield  {journal} {\bibinfo
  {journal} {Nucl. Phys. A}\ }\textbf {\bibinfo {volume} {484}},\ \bibinfo
  {pages} {557} (\bibinfo {year} {1988})}\BibitemShut {NoStop}%
\bibitem [{\citenamefont {Ashery}\ \emph {et~al.}(1981)\citenamefont {Ashery},
  \citenamefont {Navon}, \citenamefont {Azuelos}, \citenamefont {Walter},
  \citenamefont {Pfeiffer},\ and\ \citenamefont {Schleputz}}]{Ashery:1981tq}%
  \BibitemOpen
  \bibfield  {author} {\bibinfo {author} {\bibfnamefont {D.}~\bibnamefont
  {Ashery}}, \bibinfo {author} {\bibfnamefont {I.}~\bibnamefont {Navon}},
  \bibinfo {author} {\bibfnamefont {G.}~\bibnamefont {Azuelos}}, \bibinfo
  {author} {\bibfnamefont {H.}~\bibnamefont {Walter}}, \bibinfo {author}
  {\bibfnamefont {H.}~\bibnamefont {Pfeiffer}}, \ and\ \bibinfo {author}
  {\bibfnamefont {F.}~\bibnamefont {Schleputz}},\ }\href {\doibase
  10.1103/PhysRevC.23.2173} {\bibfield  {journal} {\bibinfo  {journal} {Phys.
  Rev. C}\ }\textbf {\bibinfo {volume} {23}},\ \bibinfo {pages} {2173}
  (\bibinfo {year} {1981})}\BibitemShut {NoStop}%
\bibitem [{\citenamefont {Ashery}\ \emph {et~al.}(1984)\citenamefont {Ashery}
  \emph {et~al.}}]{Ashery:1984ne}%
  \BibitemOpen
  \bibfield  {author} {\bibinfo {author} {\bibfnamefont {D.}~\bibnamefont
  {Ashery}} \emph {et~al.},\ }\href {\doibase 10.1103/PhysRevC.30.946}
  {\bibfield  {journal} {\bibinfo  {journal} {Phys. Rev. C}\ }\textbf {\bibinfo
  {volume} {30}},\ \bibinfo {pages} {946} (\bibinfo {year} {1984})}\BibitemShut
  {NoStop}%
\bibitem [{\citenamefont {Jones}\ \emph {et~al.}(1993)\citenamefont {Jones}
  \emph {et~al.}}]{Jones:1993ps}%
  \BibitemOpen
  \bibfield  {author} {\bibinfo {author} {\bibfnamefont {M.}~\bibnamefont
  {Jones}} \emph {et~al.},\ }\href {\doibase 10.1103/PhysRevC.48.2800}
  {\bibfield  {journal} {\bibinfo  {journal} {Phys. Rev. C}\ }\textbf {\bibinfo
  {volume} {48}},\ \bibinfo {pages} {2800} (\bibinfo {year}
  {1993})}\BibitemShut {NoStop}%
\bibitem [{\citenamefont {Benhar}\ \emph {et~al.}(1994)\citenamefont {Benhar},
  \citenamefont {Fabrocini}, \citenamefont {Fantoni},\ and\ \citenamefont
  {Sick}}]{Benhar:1994hw}%
  \BibitemOpen
  \bibfield  {author} {\bibinfo {author} {\bibfnamefont {O.}~\bibnamefont
  {Benhar}}, \bibinfo {author} {\bibfnamefont {A.}~\bibnamefont {Fabrocini}},
  \bibinfo {author} {\bibfnamefont {S.}~\bibnamefont {Fantoni}}, \ and\
  \bibinfo {author} {\bibfnamefont {I.}~\bibnamefont {Sick}},\ }\href {\doibase
  10.1016/0375-9474(94)90920-2} {\bibfield  {journal} {\bibinfo  {journal}
  {Nucl. Phys. A}\ }\textbf {\bibinfo {volume} {579}},\ \bibinfo {pages} {493}
  (\bibinfo {year} {1994})}\BibitemShut {NoStop}%
\bibitem [{\citenamefont {Rocco}\ \emph {et~al.}(2016)\citenamefont {Rocco},
  \citenamefont {Lovato},\ and\ \citenamefont {Benhar}}]{Rocco:2015cil}%
  \BibitemOpen
  \bibfield  {author} {\bibinfo {author} {\bibfnamefont {N.}~\bibnamefont
  {Rocco}}, \bibinfo {author} {\bibfnamefont {A.}~\bibnamefont {Lovato}}, \
  and\ \bibinfo {author} {\bibfnamefont {O.}~\bibnamefont {Benhar}},\ }\href
  {\doibase 10.1103/PhysRevLett.116.192501} {\bibfield  {journal} {\bibinfo
  {journal} {Phys. Rev. Lett.}\ }\textbf {\bibinfo {volume} {116}},\ \bibinfo
  {pages} {192501} (\bibinfo {year} {2016})},\ \Eprint
  {http://arxiv.org/abs/1512.07426} {arXiv:1512.07426 [nucl-th]} \BibitemShut
  {NoStop}%
\bibitem [{\citenamefont {Benhar}\ \emph {et~al.}(2008)\citenamefont {Benhar},
  \citenamefont {day},\ and\ \citenamefont {Sick}}]{Benhar:2006wy}%
  \BibitemOpen
  \bibfield  {author} {\bibinfo {author} {\bibfnamefont {O.}~\bibnamefont
  {Benhar}}, \bibinfo {author} {\bibfnamefont {D.}~\bibnamefont {day}}, \ and\
  \bibinfo {author} {\bibfnamefont {I.}~\bibnamefont {Sick}},\ }\href {\doibase
  10.1103/RevModPhys.80.189} {\bibfield  {journal} {\bibinfo  {journal} {Rev.
  Mod. Phys.}\ }\textbf {\bibinfo {volume} {80}},\ \bibinfo {pages} {189}
  (\bibinfo {year} {2008})},\ \Eprint {http://arxiv.org/abs/nucl-ex/0603029}
  {arXiv:nucl-ex/0603029} \BibitemShut {NoStop}%
\bibitem [{\citenamefont {Cruz-Torres}\ \emph {et~al.}(2020)\citenamefont
  {Cruz-Torres} \emph {et~al.}}]{Cruz-Torres:2020uke}%
  \BibitemOpen
  \bibfield  {author} {\bibinfo {author} {\bibfnamefont {R.}~\bibnamefont
  {Cruz-Torres}} \emph {et~al.} (\bibinfo {collaboration} {Jefferson Lab Hall A
  Tritium}),\ }\href {\doibase 10.1103/PhysRevLett.124.212501} {\bibfield
  {journal} {\bibinfo  {journal} {Phys. Rev. Lett.}\ }\textbf {\bibinfo
  {volume} {124}},\ \bibinfo {pages} {212501} (\bibinfo {year} {2020})},\
  \Eprint {http://arxiv.org/abs/2001.07230} {arXiv:2001.07230 [nucl-ex]}
  \BibitemShut {NoStop}%
\bibitem [{\citenamefont {Duer}\ \emph {et~al.}(2019)\citenamefont {Duer} \emph
  {et~al.}}]{Duer:2018sxh}%
  \BibitemOpen
  \bibfield  {author} {\bibinfo {author} {\bibfnamefont {M.}~\bibnamefont
  {Duer}} \emph {et~al.} (\bibinfo {collaboration} {CLAS}),\ }\href {\doibase
  10.1103/PhysRevLett.122.172502} {\bibfield  {journal} {\bibinfo  {journal}
  {Phys. Rev. Lett.}\ }\textbf {\bibinfo {volume} {122}},\ \bibinfo {pages}
  {172502} (\bibinfo {year} {2019})},\ \Eprint
  {http://arxiv.org/abs/1810.05343} {arXiv:1810.05343 [nucl-ex]} \BibitemShut
  {NoStop}%
\end{thebibliography}%
